\documentclass[11pt]{article}

\usepackage{booktabs}
\usepackage[margin=15pt,font=small,labelfont={bf,sf},justification=justified]{caption}
\usepackage[superscript,biblabel]{cite}
\usepackage{color}
\usepackage{float}
\usepackage{graphicx}
\usepackage{lscape}
\usepackage[bbgreekl]{mathbbol}
\usepackage{mathtools}
\usepackage{multibib}
\usepackage{multirow}
\usepackage{tabularx}
\usepackage{titleref}
\usepackage{url}
\usepackage{subcaption} 

\newif\ifbembo
\newif\ifcharter
\newif\iferewhon
\newif\iflibertine
\newif\iflibertinealt
\newif\ifpalantino
\newif\iftimesnewroman

\bembofalse
\chartertrue
\erewhonfalse
\libertinefalse
\libertinealtfalse
\palantinofalse
\timesnewromanfalse

\ifbembo
\usepackage[p,osf]{ETbb} 
\usepackage[scaled=.95,type1]{cabin} 
\usepackage[varqu,varl]{zi4}
\usepackage[T1]{fontenc}
\usepackage[libertine,vvarbb]{newtxmath}
\usepackage[cal=boondoxo,bb=boondox]{mathalfa}
\fi

\ifcharter
\usepackage[scaled=.98,p]{XCharter}
\usepackage[scaled=1.04,varqu,varl]{inconsolata}
\usepackage[type1]{cabin}
\usepackage[uprightscript,xcharter,vvarbb,scaled=1.05]{newtxmath}
\usepackage[cal=euler,scr=boondoxo,bb=boondox]{mathalpha}  
\fi

\iferewhon
\usepackage[p,osf,scaled=.98,space]{erewhon} 
\usepackage[varqu,varl]{inconsolata} 
\usepackage[type1,scaled=.95]{cabin} 
\usepackage[utopia,vvarbb]{newtxmath}
\fi

\iflibertine
\usepackage[sb]{libertinus}
\usepackage[T1]{fontenc}
\usepackage{textcomp}
\usepackage[varqu,varl]{zi4}
\usepackage{libertinust1math} 
\usepackage[scr=boondoxo,bb=boondox]{mathalpha} 
\fi

\iflibertinealt
\usepackage[sb]{libertinus}
\usepackage[T1]{fontenc}
\usepackage{textcomp}
\usepackage{libertinust1math} 
\usepackage[scr=boondoxo,bb=boondox]{mathalpha} 
\fi

\ifpalantino
\usepackage[largesc]{newpxtext}
\usepackage[vvarbb]{newpxmath}
\fi

\iftimesnewroman
\usepackage{newtxtext} 
\usepackage[scaled=0.95]{inconsolata} 
\usepackage[vvarbb]{newtxmath} 
\fi

\usepackage[T1]{fontenc}
\usepackage[final,protrusion=true,expansion=true]{microtype}

\usepackage{titling}
\usepackage{authblk} 
\pretitle{\begin{center}\bfseries\sffamily\LARGE}
	\posttitle{\end{center}}
\usepackage{abstract}

\usepackage{sectsty} 
\allsectionsfont{\sffamily}

\usepackage[left=1in,right=1in,top=1in,bottom=1in,includefoot,heightrounded]{geometry}

\newcites{supp}{SUPPLEMENTAL REFERENCES}

\makeatletter
\patchcmd{\LS@rot}{90}{-90}{}{}
\patchcmd{\endlandscape}{90}{-90}{}{}
\makeatother

\newcommand{\QS}[1]{{\color{black}#1}}
\renewcommand{\vec}[1]{\ensuremath\boldsymbol{#1}}

\newcommand{\bF}{\vec F}

\newcommand{\bP}{\vec P}

\newcommand{\bS}{\vec S}

\newcommand{\bV}{\vec V}

\newcommand{\blf}{\vec f}
\newcommand{\bn}{\vec n}

\newcommand{\bu}{\vec u}
\newcommand{\bU}{\vec U}
\newcommand{\bX}{\vec{X}}

\newcommand{\bx}{\vec{x}}

\newcommand{\bbf}{\vec f}

\newcommand{\V}{\mathcal V}

\DeclareGraphicsExtensions{.pdf,.eps,.ps,.eps.gz,.ps.gz,.eps.Y}

\def \mfac{M_{\text{fac}}}
\def \CL{C_{\text{L}}}
\def \CD{C_{\text{D}}}

\def \Re{\textrm{Re}}

\newcommand{\bxi}{\vec{\xi}}
\newcommand{\bchi}{\vec{\chi}}
\newcommand{\bphi}{\vec{\phi}}

\def \sigmaf{\bbsigma_\text{f}}

\newcommand{\norm}[1]{\left\lVert #1 \right\lVert}

\newcommand{\compositenorm}[1]{{\left\vert\kern-0.25ex\left\vert\kern-0.25ex\left\vert #1
		\right\vert\kern-0.25ex\right\vert\kern-0.25ex\right\vert}}

\newcommand{\jumpinverse}[1]{\llbracket  #1\rrbracket}

\date{\vspace{-5ex}}

\begin{document}
	\title{Improving the robustness of the immersed interface method through regularized velocity reconstruction}
	\author[1]{Qi Sun}
	\author[1]{Ebrahim M.~Kolahdouz}
	\author[1--6]{Boyce E.~Griffith}
	\affil[1]{Department of Mathematics, University of North Carolina, Chapel Hill, NC, USA}
	\affil[2]{Department of Biomedical Engineering, University of North Carolina, Chapel Hill, NC, USA}
	\affil[3]{Department of Applied Physical Sciences, University of North Carolina, Chapel Hill, NC, USA}
	\affil[4]{Carolina Center for Interdisciplinary Applied Mathematics, University of North Carolina, Chapel Hill, NC, USA}
	\affil[5]{Computational Medicine Program, University of North Carolina School of Medicine, Chapel Hill, NC, USA}
	\affil[6]{McAllister Heart Institute, University of North Carolina School of Medicine, Chapel Hill, NC, USA\vspace{\baselineskip}}
	\maketitle
	
\begin{abstract}
	Robust, broadly applicable fluid-structure interaction (FSI)  algorithms remain a challenge for computational mechanics. Efforts in this area are driven by the need to enhance predictive accuracy and efficiency in FSI simulations, align with experimental observations, and unravel complex multiscale and multiphysics phenomena, while addressing challenges in developing more robust and efficient methodologies. In previous work, we introduced an immersed interface method (IIM) for discrete surfaces and an extension based on an immersed Lagrangia-Eulerian (ILE) coupling strategy for modeling FSI involving complex geometries. The ability of the method to sharply resolve stress discontinuities induced by singular immersed boundary forces in the presence of low-regularity geometrical representations {\color{black}enables it to model complex three-dimensional geometries in diverse engineering applications}. Although the IIM we previously introduced offers many advantages {\color{black} compared to other FSI algorithms}, it also imposes a restrictive mesh factor ratio, requiring the surface mesh to be coarser than the background fluid grid to ensure stability. This is because if the mesh factor ratio constraint is not satisfied, parts of the structure motion are not controlled by the discrete FSI \QS{coupling operators}. This constraint can significantly increase computational costs, particularly in applications involving multiscale geometries with highly localized complexity or fine-scale features. To address this limitation, we devise a stabilization strategy for the velocity interpolation operator inspired by Tikhonov regularization.  \QS{The effectiveness of the stabilization scheme is evaluated using benchmark problems with stationary interfaces and FSI models involving both rigid-body dynamics and elastodynamic structures models.} This study demonstrates that using a stabilized velocity interpolation operator in the IIM enables a broader range of structure-to-fluid grid-size ratios without compromising accuracy or altering the flow dynamics. {\color{black}Our approach} significantly broadens the applicability of the method to real-world FSI problems involving complex geometries and dynamic conditions, offering a robust and practical solution.
\end{abstract}

\section{Introduction}
\label{sec:intro}

Engineering and scientific computing communities continue to be deeply invested in developing robust and efficient computational algorithms for fluid-structure interaction (FSI) problems. This continued focus stems from the growing demand for simulations in modern engineering applications and nature-inspired scientific investigations, which often involve intricate geometric and material nonlinearities alongside complex multiscale and multiphysics interactions~\cite{van_2020}. Additional demands arise from real-world scenarios, particularly systems such as medical devices, spacecraft, and nuclear reactors, \QS{in which} high-fidelity simulations, rigorously verified and validated against experimental results, are critical~\cite{jne2010005,Bause2025SystematicHI}. Over the decades, many computational {\color{black}methods for FSI} have been developed, grounded in various algorithmic paradigms.
One major classification is based on spatial discretization and grid configuration, categorizing methods into two main types: body-fitted and non-body-fitted approaches. In body-fitted methods, the computational grids of the fluid and solid domains align at the physical boundaries, whereas non-body-fitted methods typically use a fixed grid for the fluid domain that does not conform to the solid mesh.
Among most well-known body-fitted methods are arbitrary Lagrangian Eulerian (ALE) methods~\cite{Donea2004}. However, ALE methods often involve significant computational overhead and complex remeshing or mesh morphing procedures, {\color{black}particularly} when dealing with large interface displacements or deformations. In contrast, non-body-fitted methods allow the interface to cut through the elements of a fixed background grid. The main advantage of this group of methods is the relative ease of handling cases with time-dependent domains, implicitly defined domains, and domains with strong geometric deformations.  The immersed finite element method~\cite{ADJERID2015170} and CutFEM/TraceFEM~\cite{OlshanskiiQiQuaini2021,OlshanskiiQiQuaini2022} are examples of unfitted approaches. One complex aspect of these methods is the need for tailored stabilization. A widely used approach in this category is the immersed boundary (IB) method introduced by Peskin~\cite{PESKIN1972252,Peskin_2002}. The IB method provides a means to incorporate the effects of a solid boundary into the fluid equations through the localization of singular forces at the interface. Typical discretizations employ a fixed Cartesian grid for the fluid domain and represent the structure using discrete Lagrangian points. FSI is mediated through integral transforms with regularized Dirac delta function kernels, which the IB method uses to connect Lagrangian and Eulerian variables. 

 IB-type approaches using regularized delta functions are appealing for their ease of implementation and flexibility in handling large deformations. The major limitation of regularized methods lies in the need to resolve the smoothing region with an adequate number of elements, which often leads to lower accuracy and convergence rates near the interface~\cite{GRIFFITH200575}. Even with a refined grid, spurious feedback forces and pressure oscillations may be still present, generating undesirable internal flows within the structure. Other challenges include reliably calculating hemodynamic stresses such as wall shear stress, and handling multiple immersed boundaries in close proximity~\cite{FAI2018319}. 

The immersed interface method (IIM)~\cite{leveque1994immersed} is an alternative to the IB method that was originally developed to improve the accuracy of the IB method. {\color{black} In this approach, correction terms arising from the interface conditions must be incorporated into the associated fluid equations to resolve stress discontinuities and interfacial boundary conditions sharply and to achieve higher accuracy systematically.} To sharply resolve stress discontinuities and interfacial boundary conditions and achieve higher accuracy, correction terms arising from the interface conditions must be incorporated into the associated fluid equations. LeVeque and Li applied the concept of jump conditions to develop solutions for elliptic equations with discontinuous coefficients or singular forces~\cite{leveque1994immersed}. The IIM was then used to solve Stokes and Navier-Stokes~\cite{LeVeque97,LeVequeRandall2003}. In a more recent work by some of us, we introduced an IIM for complex geometries described by discrete surfaces, making the IIM more accessible for real-world engineering simulations involving experimental or clinical image data~\cite{KOLAHDOUZ2020108854}. In subsequent work, we designed an immersed Lagrangian Eulerian (ILE) method for FSI of both rigid and flexible structures with a fluid solver and coupling strategy that base on our IIM for discrete surfaces~\cite{KOLAHDOUZ2021110442, KOLAHDOUZ2023112174}.

Similar to the IB method, the IIM uses a combination of Eulerian and Lagrangian variables. These variables are coupled through interaction equations, with the interface jump conditions playing a fundamental role~\cite{xusheng06,Lai2001ARO}. In the IIM, the Eulerian variables are defined on a fixed Cartesian mesh, while the Lagrangian variables are defined on a curvilinear mesh that moves freely through the fixed Cartesian mesh. In our previous work~\cite{KOLAHDOUZ2020108854,KOLAHDOUZ2021110442, KOLAHDOUZ2023112174}, we had to ensure that the numerical discretization satisfied a restriction on the mesh factor ratio, $\mfac = h_{\text{L}} / h_{\text{E}} > 1$, in which $h_{\text{L}}$ is the local Lagrangian element size and $h_{\text{E}}$ is the local Eulerian grid spacing. For a relatively uniform triangulation of the immersed boundary, $h_{\text{L}}$ typically is 
the target mesh size specified during the mesh generation process. For complex geometries, {\color{black}it can be difficult or impossible to control precisely} the minimum and maximum element sizes during the meshing process, or in situations where a specific mesh refinement rule is imposed, the local  element size $h_{\text{L}}$ may be significantly smaller than the target element size.
{\color{black}Preliminary studies and numerical simulations of the methodology suggest the mesh factor ratio must globally satisfy the condition $\mfac>1$.} 
{\color{black}This is based on results from} previous numerical experiments indicate that smaller $\mfac$ values often lead to numerical instability.
{\color{black}Satisfying this} restriction therefore often requires refining of the Eulerian grid. Because the fluid solver is typically the most  {\color{black}expensive operation in an IIM computation}, 
satisfying this constraint can substantially increase the overall computational cost, especially for interface geometries with dense local structures and large deformations.

The key contribution of this paper is that it proposes a stabilization scheme based on Tikhonov regularization~\cite{tikhonov1977solutions} to relax the mesh factor ratio constraint $\mfac > 1$ {\color{black}so that the method can support a broad range of $\mfac$ values.} In the IIM discretization of our FSI model, interfacial velocities are interpolated from the background Cartesian grid to the interface using a velocity interpolation operator. To satisfy the no-slip and no-penetration interface conditions, we use this interfacial velocity to update the interface position. The interfacial force is then {\color{black}computed} by a motion discrepancy penalty method~\cite{GOLDSTEIN1993354}. {\color{black}Interfacial forces are transmitted from the boundary back to the Cartesian grid through a force spreading operator.} If the mesh factor ratio constraint is not satisfied, the force spreading operator {\color{black}will have a nontrivial null space}. As a result, part of the interface motion and velocity have no impact on the forces that appear in the fluid equations. This creates instability because part of the surface motion is not controlled by the discrete FSI system. The main idea of {\color{black}our stabilization approach} is to regularize the interfacial velocity along, but not across, the boundary, so that the motion and velocity of the interface are fully controlled by the discrete system. \QS{To demonstrate the effectiveness of our stabilization scheme, we consider problems with stationary interface configurations, as well as FSI models involving both rigid-body dynamics and elastodynamic structures models. Numerical tests demonstrate that our stabilized formulation maintains accuracy comparable to our previously proposed IIM while allowing for much smaller or highly  disparate structure-to-fluid grid-size ratios previously considered infeasible. }

\section{Mathematical formulation}
\label{sec:formulation}

\begin{figure}[b!!]
	\centering
	\includegraphics[width=0.3\textwidth]{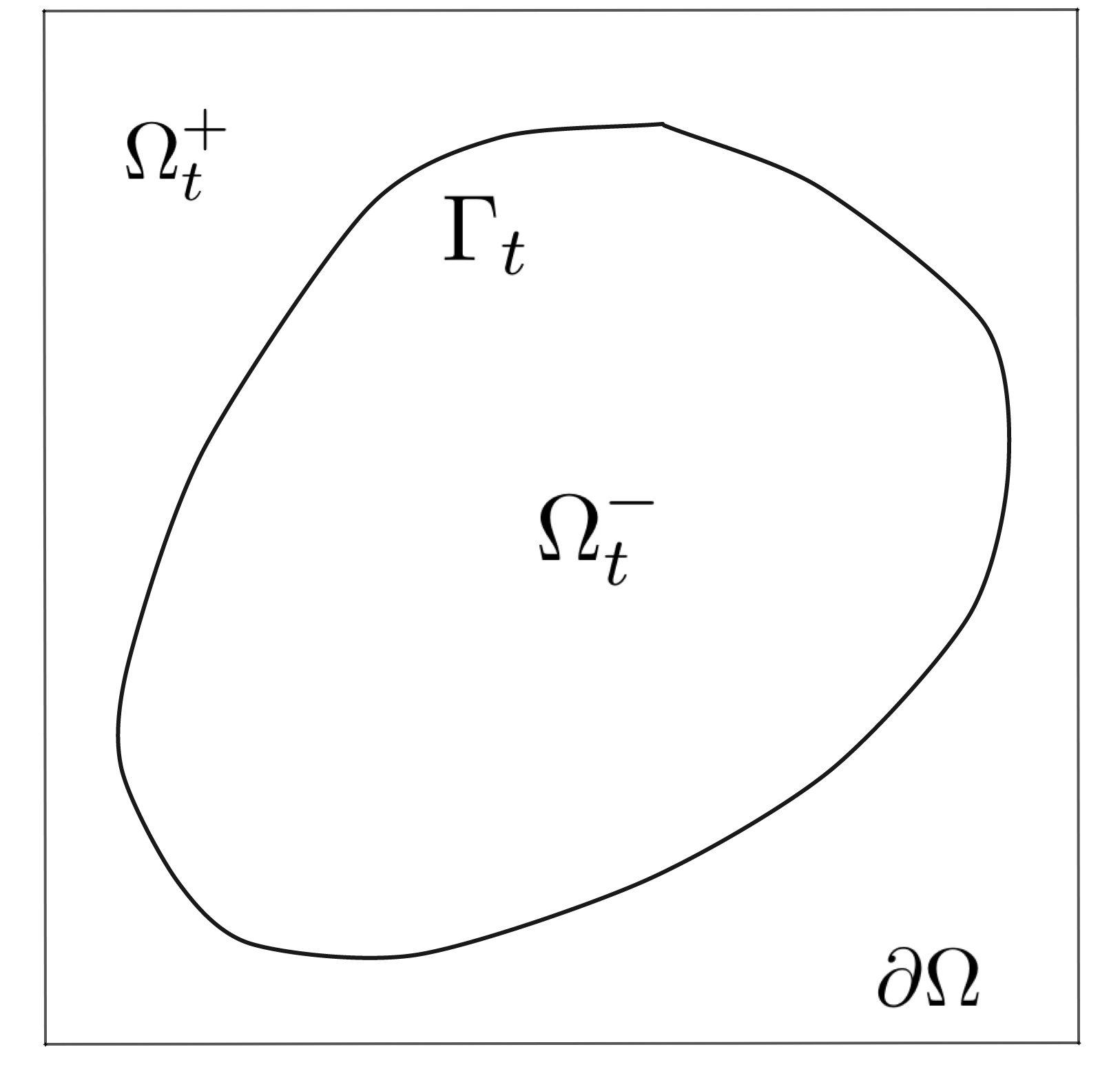}
	\caption{Fluid domains and interface in $\mathbb{R}^2$.}
	\label{fig:Geom}
\end{figure}

We consider a domain $\Omega\subset\mathbb{R}^d$, {\color{black}$d=2$ or $3$}, that is divided into an external subdomain $\Omega^+_t$ and an internal subdomain $\Omega^-_t=\Omega\setminus\overline{\Omega^+_t}$, each parameterized by time $t$. The immersed interface is $\Gamma_t=\overline{\Omega^+_t}\cap\overline{\Omega^-_t}$; see Fig.~\ref{fig:Geom}.
We describe quantities associated with the interface using reference coordinates  $\bX\in\Gamma_0$  attached to the interface at time $t=0$. $\bchi(\bX, t)$ is the interface position, and the interface velocity is  $\bU(\bX, t) = \frac{\partial \bchi}{\partial t}(\bX, t)$. We assume that the fluid mass density $\rho$ and dynamic viscosity $\mu$ are uniform throughout $\Omega$. The Cauchy stress tensor  $\sigmaf(\bx,t)$ takes the form 
$\sigmaf=-p \mathbb{I} +\mu( \nabla \bu + (\nabla\bu)^{\transp})$. 
The fluid-structure interaction system is described by:
	\begin{align}
		\rho\frac{{\mathrm D}\bu}{\mathrm {Dt}}(\bx,t)&= \nabla \cdot \sigmaf(\bx,t),  & \bx\in \Omega, \label{eqn:NS_momentum}\\
		\nabla\cdot\bu(\bx,t)&=0, & \bx\in \Omega,\label{eqn:NS_incompressible}\\
		\jumpinverse{\sigmaf(\bchi(\bX,t),t),t)\bn(\bchi(\bX,t))}&=-j^{-1}(\bX,t)\bF(\bX,t),& \bX\in \Gamma_0,\label{eqn:NS_Jump_stress}\\
		\frac{\partial \bchi}{\partial t}(\bX,t)&=\bu(\bchi(\bX,t),t),&\bX\in\Gamma_0,\label{eqn:bcd}\\
		\bF(\bX,t)&=\kappa(\bxi(\bX,t)-\bchi(\bX,t)),&\bX\in\Gamma_0,\label{eqn:feedback_force}
	\end{align}
in which  $\bu(\bx,t)$ and $p(\bx,t)$ are the fluid velocity and pressure respectively. $j(\bX,t)$ is the surface Jacobian determinant that converts the {\color{black}interfacial} force density from force per unit area in the current configuration to force per unit area in the reference configuration. $\jumpinverse{\cdot}$ {\color{black} indicates} the jump in the bracketed quantity across the interface along the normal from the exterior to the interior region. Eq.~(\ref{eqn:bcd}) corresponds to the no-slip, no-penetration conditions. To simplify the description of our approach, we simplify the fluid-structure model by assuming that the physical motion and position of the material interface $\Gamma_0$ are prescribed functions of time. Specifically, we  prescribe the physical position of a material point $\bX$ at time $t$ as $\bxi(\bX, t)$, and the velocity as $\bV(\bX, t) = \frac{\partial \bxi}{\partial t}(\bX, t)$. Note that this setting can be easily extended to a rigid-body or flexible-body fluid-structure interaction model, as in our previous work~\cite{KOLAHDOUZ2023112174,KOLAHDOUZ2021110442}, in which the motion $\bxi(\bX,t)$ of the material interface $\Gamma_0$ is governed by the equations of the body motion.  In Eq.~(\ref{eqn:feedback_force}), we adopt a penalty force formulation similar to that proposed by Goldstein et al.~\cite{GOLDSTEIN1993354} and also used in our previous studies involving the IIM for discrete surfaces~\cite{KOLAHDOUZ2020108854,KOLAHDOUZ2021110442, KOLAHDOUZ2023112174}. $\bF(\bX,t)$ is the interfacial force along the interface that{\color{black}, in a penalty method, corresponds to an approximate Lagrange multiplier force}.  Here, $\kappa>0$  is a spring stiffness constant. In principle, we want to choose $\kappa$ {\color{black} as large as possible to minimize the deviation of the interface}. In the stationary interface case, Eq.~(\ref{eqn:feedback_force}) becomes $
\bF(\bX,t)=\kappa\left( \bX-\bchi(\bX,t) \right).$

\section{Numerical Discretization}
\label{sec:numerical}

This section outlines the numerical method used for discretizing the fluid-structure model.  To simplify the notation, the numerical scheme is presented in two spatial dimensions. The extension of the method to three spatial dimensions is straightforward~\cite{KOLAHDOUZ2020108854}.

\subsection{Fluid-structure coupling}
We discretize the incompressible Navier-Stokes equations using an adaptively refined staggered grid discretization, as in our previous work~\cite{GRIFFITH20097565,Boyce052011}, which approximates the pressure $p$ at cell centers and the velocity $\bu$ and forcing terms $\blf$ at the edges (in two dimensions) or faces (in three dimensions) of the {\color{black}Cartesian }grid cells. \QS{For} the fluid-structure coupling scheme, we begin by considering a family of triangulations $\{\textit{T}_h \}$ for $\Gamma_0$. The domain formed by $\textit{T}_h $ is denoted by $\Gamma_{h,0}=\cup_{\textit{T}\in \{\textit{T}_h \}}\overline{\textit{T}}$. We use a finite element space to represent interfacial Lagrangian variables. We denote the finite element space by $\V_h$. A spatial discrete analogue of the Eqs.~(\ref{eqn:NS_Jump_stress})--(\ref{eqn:feedback_force}) is:
\begin{align}
	\bbf&=\bS_h[\bchi]\bF,\label{spatial_eqn:force_spreading}\\
	\frac{d\bchi}{d t}&=\bU=\vec{J}_h^{\epsilon}[\bchi,\bF](\bu),\label{spatial_eqn:interpolation}\\
	\bF&=\kappa(\bxi-\bchi), \label{eqn:dc_spring_force}
\end{align}
in which $\bF,\bU,\bV,\bchi$ and $\bxi\in [\V_h]^d$. We adopt and extend the IIM coupling scheme based on our earlier work~\cite{KOLAHDOUZ2020108854}.  
The force spreading operator $\vec{S}_h$ in Eq.~(\ref{spatial_eqn:force_spreading}) and unregularized velocity interpolation operator $\vec{J}_h$ in Eq.~(\ref{spatial_eqn:interpolation}) \QS{are} constructed based on the interface jump conditions. For more technical details \QS{on the form of} $\vec{S}_h$, we refer to Kolahdouz et al.~\cite{KOLAHDOUZ2020108854}, \QS{who discuss} the construction of $\vec{S}_h$ through introducing correction terms in the discretization of the momentum equation. For the velocity interpolation operator $\vec{J}_h$, we use the method detailed by Tan et al.~\cite{tan2009}. The stabilized velocity interpolation operator is defined as   $\vec{J}_h^{\epsilon}=\vec{P}_h^{\epsilon}\vec{J}_h$, in which $\vec{P}_h^{\epsilon}$ represents a modified $L^2$ Tikhonov regularization operator~\cite{tikhonov1977solutions}. More details on the construction of $\vec{P}_h^{\epsilon}$ are provided in the next subsection. Briefly, however, if  $\epsilon=0$, $\vec{P}_h^{0}$ is a standard $L^2$ projection operator, which  was used in our earlier work~\cite{KOLAHDOUZ2020108854,KOLAHDOUZ2021110442,KOLAHDOUZ2023112174}.   

\subsection{\QS{Regularized velocity interpolation}}

\begin{figure}[!!t!!]
	\centering
	\includegraphics[scale=0.15]{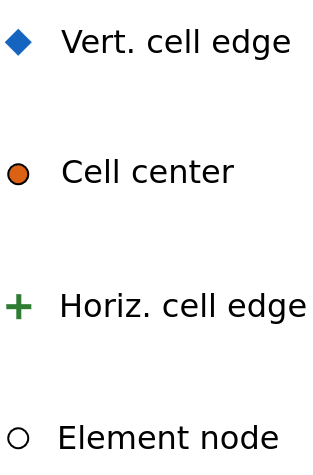}\includegraphics[scale=0.18]{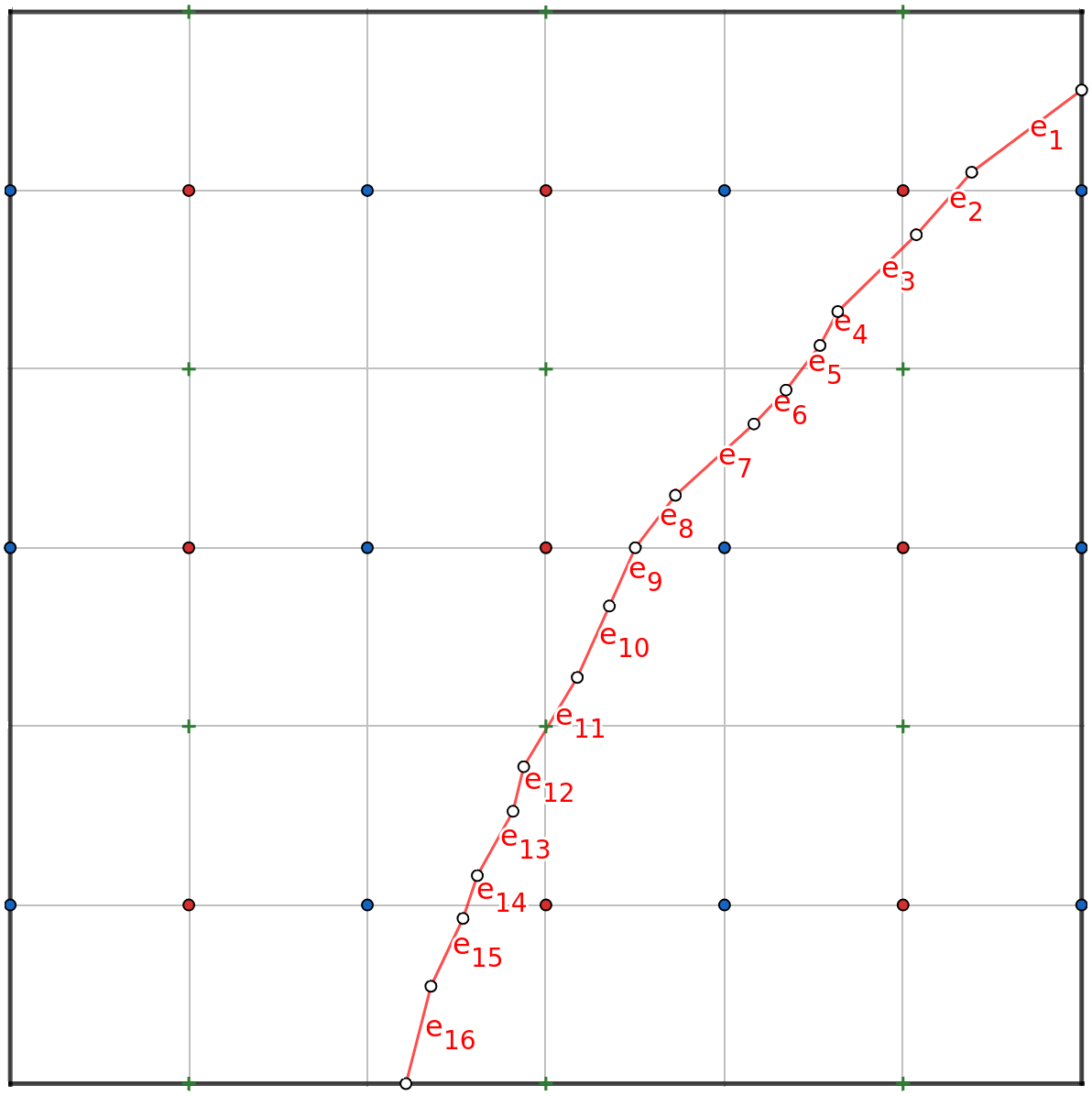}
	\caption{Illustration of a finite difference scheme for the Navier-Stokes equations and discrete interface with local $\mfac$ value approximately in the range $0.1 \leq \mfac \leq 0.25$. $\{e_i\}$ denote surface elements. Elements $e_4,e_6,e_7,e_{8},e_{10},e_{12},e_{13}$ and $e_{15}$ are not cut by any finite difference stencils.}
	\label{fig:small_Mfac}
\end{figure}

To obtain a finite element representation of the Lagrangian variables, in our previous work~\cite{KOLAHDOUZ2020108854}, we introduce a standard $L^2$ projection operator, denoted by $\vec{P}_h^0$ in Eq.~(\ref{spatial_eqn:interpolation}), to project functions from $\left[L^2(\Gamma_{h,0})\right]^d$ into $[\V_h]^d$. In the discrete problem, recall that forces are transmitted from the interface to the background Cartesian grid through the force spreading operator $\vec{S}_h$ in Eq.~(\ref{spatial_eqn:force_spreading}).  
If the mesh factor ratio constraint $\mfac > 1$ is not satisfied, the force spreading operator $\vec{S}_h$ has a nontrivial null space that includes discrete force components not associated with any element cut by a finite difference stencil; see Fig.~\ref{fig:small_Mfac}. Because $\bF$ is obtained from $\bchi$ via Eq.~(\ref{eqn:dc_spring_force}), the null space of $\vec{S}_h$ further results in the surface velocity $\bU$ and motion $\bchi$ of uncut elements not being controlled by the discrete system. 
This is physically unstable, and if it occurs in a simulation, the computation will generally also become unstable. To address this problem, we regularize the interfacial velocity $\bU$ along, but not across, the surface so that the motion $\bchi$ and velocity $\bU$ of the interface are fully controlled by the discrete system. \QS{To do so}, we use an $L^2$ Tikhonov regularization technique to define $\vec{P}_h^{\epsilon}$ in Eq.~(\ref{spatial_eqn:interpolation}). Specifically, given a function $\bphi\in[L^2(\Gamma_{h,0})]^d$, we find $\vec{P}_h^{\epsilon}(\bphi)\in [\V_h]^d$ {\color{black}that} satisfies
\begin{align}
	\int_{\Gamma_{h,0}}\vec{P}_h^{\epsilon}(\bphi)\cdot\vec{\psi}\ \text{d}A+\epsilon h^2\int_{\Gamma_{h,0}}\nabla_{\Gamma} \vec{P}_h^{\epsilon}(\bphi)\cdot\nabla_{\Gamma} \vec{\psi}\ \text{d}A=\int_{\Gamma_{h,0}}\vec{\phi}\cdot\vec{\psi}\ \text{d}A,  &\qquad \text{for any }  \vec{\psi}\in [\V_h]^d, \label{tikhnov_eqn}
\end{align}
in which $\nabla_{\Gamma_{h,0}}$ is the surface gradient, $\epsilon>0$ is a stability coefficient, and $h$ is the Cartesian grid size. {\color{black}In practice, we consider a finite element approximation of Eq.~(\ref{tikhnov_eqn}) and choose $\V_h$ to be the standard $\bP^1$ finite element space. Integration in Eq.~(\ref{tikhnov_eqn}) is computed using third order Gauss quadrature. The main idea of the modified $L^2$ Tikhonov regularization is to regularize $\bU$ along, but not across, the surface, so that the motion $\bchi$ and velocity of the interface are fully controlled by the discrete system.}  \textcolor{black}{The appropriate value for the parameter $\epsilon$ is determined empirically by identifying the minimum value that ensures the stability of the discrete system. This selection procedure is demonstrated through the numerical tests presented in Section 4.1.
}

\subsection{Time discretization} 
We use the IMEX-BDF2 scheme~\cite{Dongwang} for time discretization. To advance the system from time $t^n$ to time $t^{n+1}$, we start with $\bxi^{n+1}$ an $\bV^{n+1}$ at time $t^{n+1}$, $\bu^n, \bF^n$ and $ \bchi^{n} \text{ at time  }t^n$, and $ \bu^{n-1},\bchi^{n-1}$ , and $\bU^{n-1}\text{ at time  }t^{n-1}$,  and determine $\bchi^{n+1},  \bF^{n+1},$ and $\bbf^{n+1}$ via
\begin{align}
	\frac{\bchi^{n+1}-\bchi^{n}}{\Delta t^{n+1}}&=\beta_1\bU^{n}+\beta_2\bU^{n-1}=\beta_1\vec{J}^{\epsilon}_h[\bchi^{n},\bF^{n}](\bu^{n})+\beta_2\bU^{n-1},\\
	\bF^{n+1}&=\kappa\left(\bxi^{n+1}-\bchi^{n+1}\right),&\\
	\bbf^{n+1}&=\vec{S}_h[\bchi^{n+1}]\bF^{n+1},
\end{align}
in which $\beta_1=1+\frac{\Delta t^{n+1}}{2\Delta t^{n}}$ and $\beta_2=-\frac{\Delta t^{n+1}}{2\Delta t^{n}}$. Then we determine $\bu^{n+1}$ and $p^{n+1}$ by solving

\begin{align}
	\rho\left(\frac{\alpha_1\bu^{n+1}+\alpha_0\bu^{n}+\alpha_{-1}\bu^{n-1}}{\Delta t^{n+1}}+\vec{A}^{n+\frac{1}{2}}\right)&= \mu\vec{L}_h\bu^{n+1}-\vec{G}_h p^{n+1} +\bbf^{n+1}, \label{eqn:dc_momentum_time_discretized} \\
	\vec{D}_h\bu^{n+1}&=0, 
\end{align}
in which $\alpha_1=\left(1+\frac{\Delta t^{n+1}}{\Delta t^{n}+\Delta t^{n+1}}\right)$, $\alpha_{-1}=\frac{\Delta t^{n+1}}{\Delta t^{n}}(1+\frac{\Delta t^{n+1}}{\Delta t^{n}+\Delta t^{n+1}})$ and $ \alpha_0=-(\alpha_1+\alpha_{-1})$. The discrete divergence $\vec{D}_h$, gradient $\vec{G}_h$, and Laplace operators $\vec{L}_h$ correspond to compact, second-order accurate finite difference schemes.
$\vec{A}^{n+\frac{1}{2}}=\frac{3}{2}\vec{A}^{n}-\frac{1}{2}\vec{A}^{n-1}$ is obtained from a high-order upwind spatial discretization of the nonlinear convective term
$\bu \cdot \nabla \bu$. Discretization details are provided by Griffith~\cite{Griffith_2012}. We use the variable implicit two-step backward differentiation formula discussed by Wang et al.~\cite{Dongwang} in Eq.~(\ref{eqn:dc_momentum_time_discretized}), which requires only linear solvers for the time-dependent incompressible Stokes equations. \QS{We solve the time-dependent Stokes equations using the FGMRES solver preconditioned by a projection method-based preconditioner.~\cite{GRIFFITH20097565}} In the initial time step, a two-step predictor-corrector method is used to determine the velocity, deformation, and pressure; see Griffith and Luo~\cite{Griffith_luo_2016} for further details.

\section{Numerical Examples}

In this section, we present verification examples in both two and three spatial dimensions. The purpose of the numerical results provided here is to demonstrate the robustness of the proposed stabilized IIM scheme across a wide range of $\mfac$ values. To further assess the robustness of the stabilized IIM in more complex scenarios, we include several \QS{fluid-structure interaction test cases for both rigid and flexible immersed structures}.

Similar to our previous work~\cite{KOLAHDOUZ2021110442}, in all numerical examples, the Eulerian domain $\Omega$ is discretized using an adaptively refined grid. The Cartesian grid spacing on the finest level of refinement is  $h_{\text{finest}} = r^{-(N-1)} h_{\text{coarsest}}$, in which $h_{\text{coarsest}}$ is the grid spacing on the coarsest level, $r$ is the refinement ratio, and $N$ is the number of refinement levels.

In all tests, we choose $\V_h$ to be the standard $\bP^1$ finite element space. The penalty parameter is set as $\kappa = \frac{\kappa_0}{\Delta t^2}$, and the time step size is given by $\Delta t = c_0 h$. The constants $c_0$ and $\kappa_o$ are chosen to ensure both satisfaction of the advective CFL condition and that $\norm{\bchi(\bX,t) - \bxi(\bX,t)}_{\infty} < h$  for a given interface motion $\bxi(\bX,t)$, which may be prescribed or the result of FSI models. 
Finally, we empirically determine the smallest value of $\epsilon \geq 0$ that ensures the stability of the scheme in each case.
\begin{figure}[!!b!!]
	\centering
	\hskip -.3cm
	\includegraphics[width=0.6\textwidth]{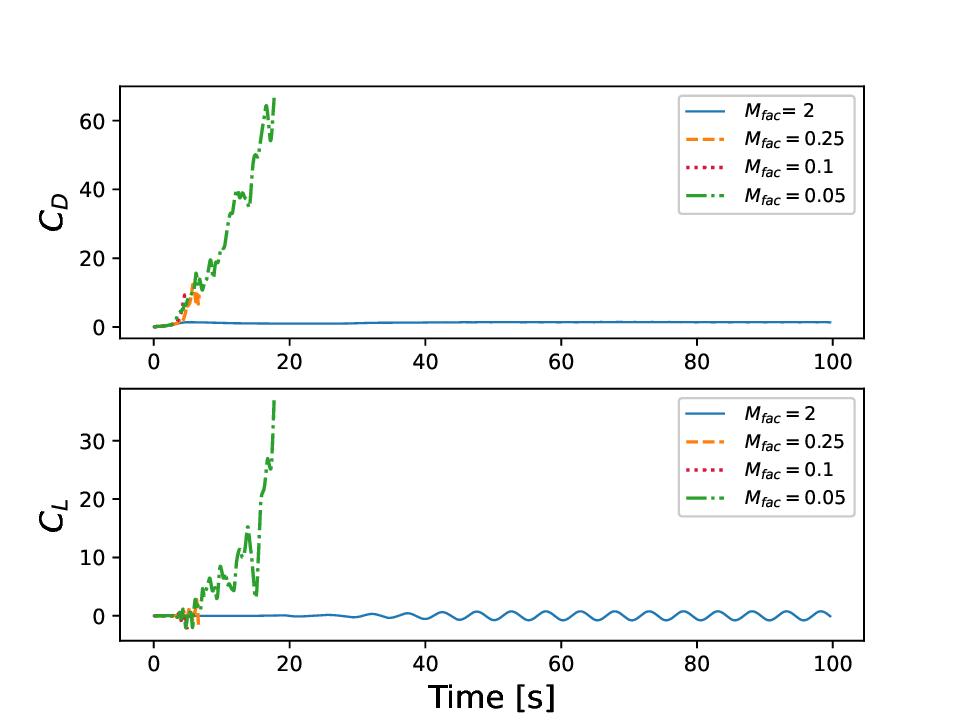}\quad
	\caption{Drag coefficient $\CD$ and lift coefficients $\CL$ over time  for two-dimensional flow past a cylinder with $\mfac=0.05,0.1,0.25,$ and $2$, generated by the unmodified IIM~\cite{KOLAHDOUZ2020108854}. {\color{black} Simulations with $\mfac < 1$ stopped because of excessive spurious interfacial motions.}}
	\label{fig:2D_flow_cylinder_no_stab}
\end{figure}

\subsection{Flow past a stationary cylinder}

This section considers a test of flow past a stationary cylinder. This is a widely used benchmark problem for testing discretization schemes. In this example, we use the configuration from the setup described in the previous work~\cite{TAIRA20072118,KOLAHDOUZ2020108854}. The computational domain is $\Omega=[-15,45]\times [-30,30]$, which is a square with length $L=60$. The immersed boundary is a stationary disc centered at the origin with a diameter 1. We impose inflow velocity boundary condition $\bu=(\frac{\tanh(t/2-2)+\tanh(2)}{1+\tanh(2)},0)$ on the boundary $(x = -15)$, zero normal traction and zero tangential velocity is imposed at the right
boundary $(x = 45)$ as an outflow condition.  Along the bottom $(x = -30)$ and top $(x = 30)$ boundaries, the
normal velocity and tangential traction are set to zero. 
We set $\rho = 1$, and use velocity $(1,0)$ as the characteristic velocity. The Reynolds number is $\Re=\frac{\rho U D}{\mu}$, which with our model parameters implies that $\mu=\frac{1}{\Re}$. We set the Reynolds number to be $\Re=200$. 

The computational domain is discretized using 6 levels of local adaptive refinement, with a refinement ratio of $r=2$. The coarsest Cartesian grid spacing is set to $h_{\text{coarsest}} = \frac{L}{256}$, and the finest is $h_{\text{finest}} = \frac{L}{2048}$. The time step size is set to $\Delta t = 0.1 h_\text{finest}$. Values of $\mfac$ ranging from 0.05 to 2 are considered.

To assess the dynamics from the numerical simulations, we compute nondimensional quantities including the drag coefficient $\CD$ and lift coefficient $\CL$ as,
\begin{align}
	(\CD,\CL)=\frac{-\int_{\Gamma_0}\bF(\bX,t)\, \mathrm{d}A }{\frac{1}{2}\rho U^2D}.
\end{align}
\begin{table}[t]
	\centering
	\begin{tabular}{|l|l|l|}
		\hline
		& $\CD$ & $\CL$ \\ \hline
		Braza et al.~\cite{Braza_Chassaing_Minh_1986}& 1.400 $\pm 0.05$  & $\pm$ 0.75  \\ \hline
		Liu et al.~\cite{LIU199835}&1.310 $\pm$ 0.049  & $\pm$ 0.69 \\ \hline
		Griffith and Luo~\cite{Griffith_luo_2016}& 1.360 $\pm$ 0.046  & $\pm$ 0.70  \\ \hline
		Xu and Wang~\cite{xusheng06}&1.420 $\pm$ 0.040&  $\pm$ 0.66\\ \hline
		IIM~\cite{KOLAHDOUZ2020108854} $\mfac=2$&1.38 $\pm$ 0.05& $\pm$ 0.77 \\ \hline
		Stabilized IIM with $\epsilon=116.5$ $\mfac=0.25$& 1.38 $\pm$ 0.05& $\pm$ 0.77\\ \hline
		Stabilized IIM with $\epsilon=116.5$ $\mfac=0.1$& 1.38 $\pm$ 0.05& $\pm$ 0.77 \\ \hline
		Stabilized IIM with $\epsilon=116.5$ $\mfac=0.05$&1.38 $\pm$ 0.05&$\pm$ 0.77 \\ \hline
		Stabilized IIM with $\epsilon=116.5$ $\mfac=0.05-0.27$&1.38 $\pm$ 0.05&$\pm$ 0.77 \\ \hline
	\end{tabular}
	\caption{Drag coefficient $\CD$ and lift coefficients $\CL$ for two-dimensional flow past a cylinder with $\Re=200$. }
	\label{tb:2D_flow_cylinder}
\end{table}

\begin{figure}[b!!]
	\centering
	\hskip -.3cm
	\includegraphics[width=0.6\textwidth]{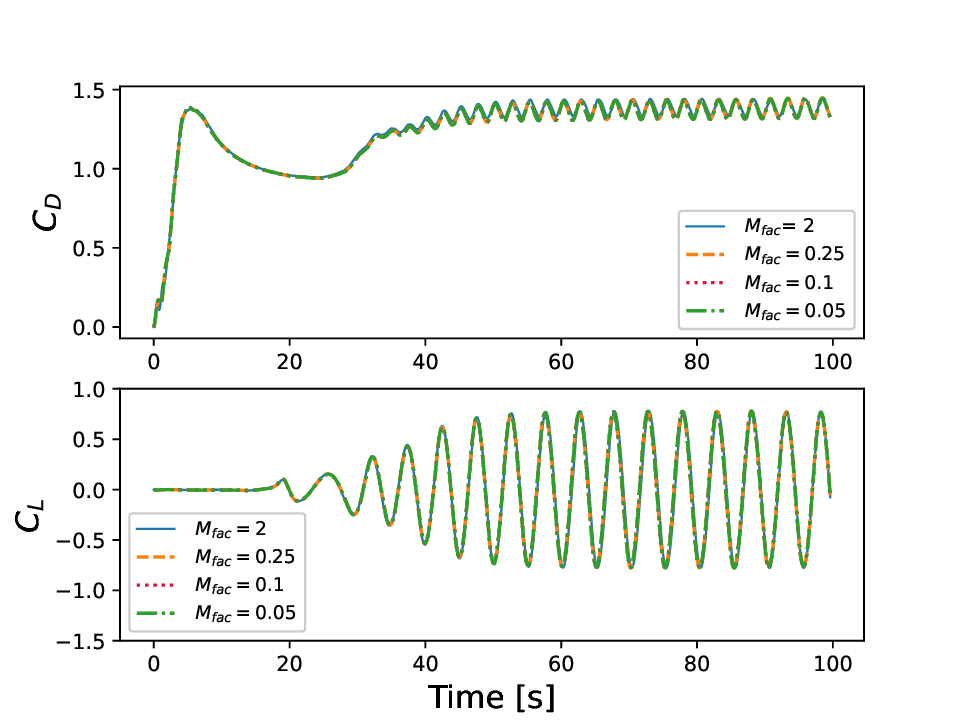}
	\caption{Drag coefficient $\CD$ and lift coefficients $\CL$ over time for two-dimensional flow past a cylinder with $\mfac=0.05,0.1,0.25$, and $2$, generated by the stabilized IIM with stability coefficient $\epsilon=116.5$.}
	\label{fig:2D_flow_cylinder_stab_0.1}
\end{figure}

\begin{figure}[t!!]
	\centering
	\includegraphics[width=.60\textwidth]{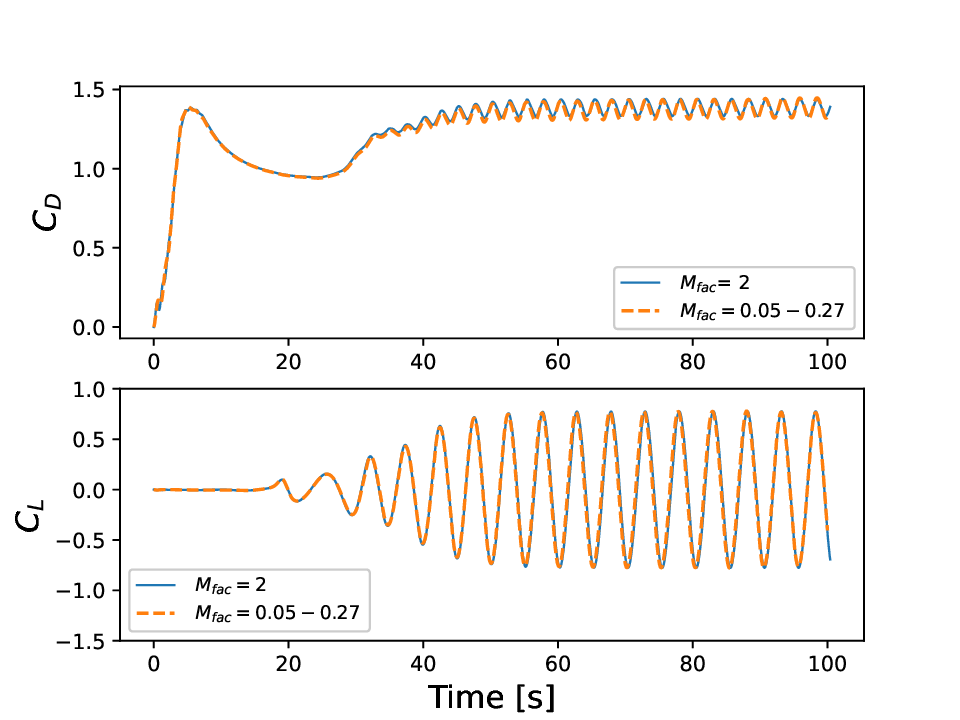}
	\caption{Drag coefficient $\CD$ and lift coefficients $\CL$ over time for two-dimensional flow past a cylinder with non-uniform interface discretization with $\mfac=0.05-0.27$  generated by the stabilized IIM with stability coefficient $\epsilon=116.5$.}
	\label{fig:non_uniform_mesh_size}
\end{figure}

To demonstrate the inherent numerical instability in the IIM that arises when small values of $\mfac$ are used, we first test the IIM without stabilization for small values of  $\mfac$, using the IIM without stabilization with $\mfac = 2$ as a reference for comparison. Fig.~\ref{fig:2D_flow_cylinder_no_stab} details the lift and drag coefficients for different values of $\mfac$ from the IIM without stabilization~\cite{KOLAHDOUZ2020108854}. Fig.~\ref{fig:2D_flow_cylinder_no_stab} demonstrates the instabilities that occur with small $\mfac$ values, ultimately leading to the simulation becoming unstable. For comparison, we perform the same experiments using the stabilized IIM for small values of $\mfac$, similarly using IIM without stabilization with $\mfac = 2$ as a reference for comparison. Fig.~\ref{fig:2D_flow_cylinder_stab_0.1} details the lift and drag coefficients for different values of $\mfac$ from the stabilized IIM proposed in the current work,  demonstrating that using the stabilized IIM with small values of $\mfac$ reproduces nearly identical lift and drag coefficient dynamics as the unmodified IIM that satisfies the $\mfac > 1$. Fig.~\ref{fig:non_uniform_mesh_size} presents the lift and drag coefficients for a nonuniform interface discretization with $25\%$ local $\mfac = 0.05$ and $75\%$ local $\mfac = 0.27$. Figs.~\ref{fig:2D_flow_cylinder_stab_0.1} and  \ref{fig:non_uniform_mesh_size}  show that our stabilization approach is robust with respect to the small mesh size ratio $\mfac$.  Table \ref{tb:2D_flow_cylinder} lists the drag coefficient $\CD$ and lift coefficient $\CL$ for $\Re = 200$, comparing values from the literature with the results obtained from stabilized IIM simulations for $\mfac = 0.05$, $0.1$, $0.25$, and for a non-uniform mesh with a range of element sizes, so that $\mfac$ is in the range $0.05$ -- $0.27$. 

\begin{figure}[b!!]
	\centering
	\includegraphics[width=.60\textwidth]{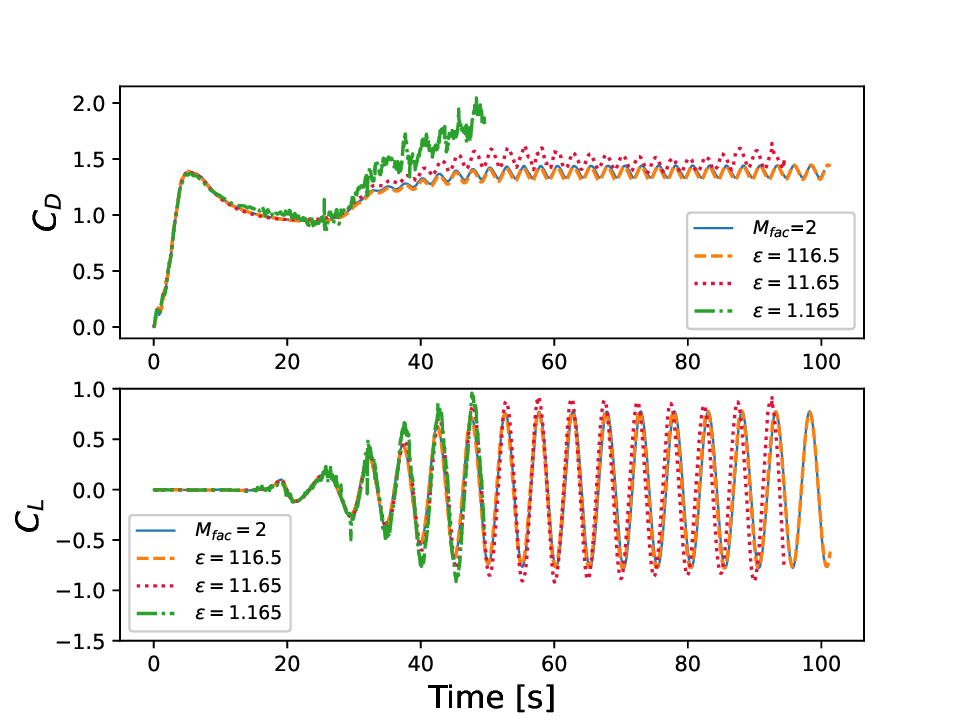}
	\caption{ Drag coefficient $\CD$ and lift coefficients $\CL$ over time for two-dimensional flow past a cylinder with $\mfac=0.25$, stabilization parameters $\epsilon=1.165,11.65,$ and $116.5$  generated by the stabilized IIM. {\color{black}Simulations with $\epsilon =1.165,11.65$ stopped because of excessive spurious interfacial motions.} }
	\label{fig:vary_eps}
\end{figure}
\begin{figure}[t!!]
	\centering
	\includegraphics[width=.6\textwidth]{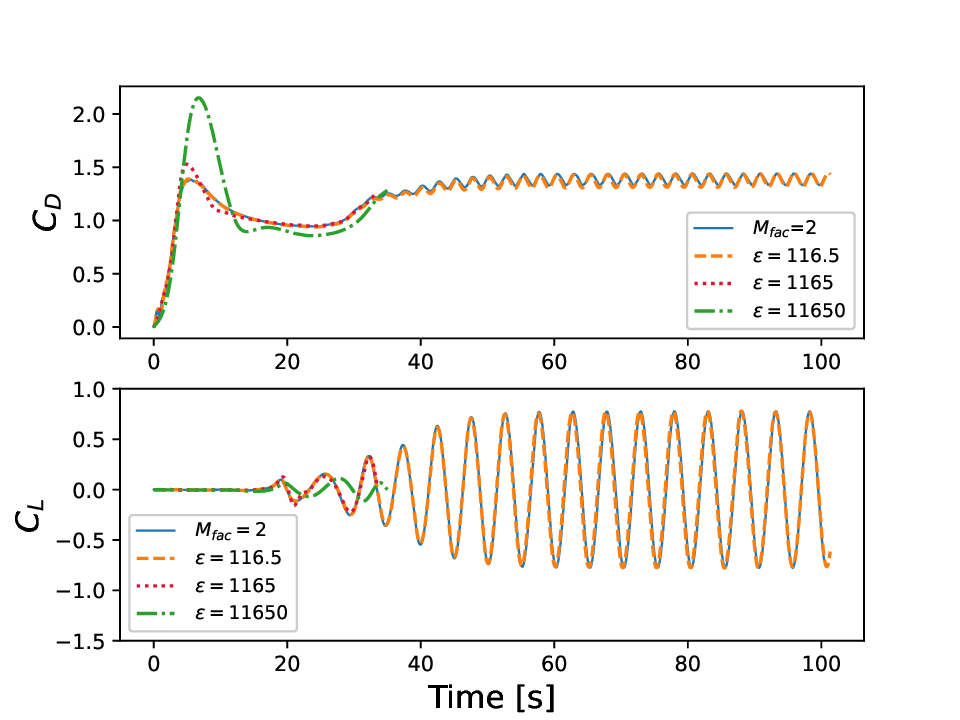}
	\caption{ Drag coefficient $\CD$ and lift coefficients $\CL$ over time for two-dimensional flow past a cylinder with $\mfac=0.25$, stabilization parameters $\epsilon=116.5,1165,$ and $11650$  generated by the stabilized IIM. {\color{black}Simulations with $\epsilon =1165,11650$ stopped because of excessive spurious interfacial motions.} }
	\label{fig:vary_eps2}
\end{figure}

To quantify the difference between dynamics of the drag and lift coefficients from the unmodified IIM with $\mfac = 2$ and the stabilized IIM with smaller values of $\mfac$, we employ the averaged relative discrepancy and averaged relative percent
difference (RPD) as metrics to evaluate the discrepancy; see Table \ref{table: MSE_vary_Mfac}. Our stabilized IIM simulations for a smaller value of $\mfac$ yield excellent quantitative agreement with the unmodified IIM at $\mfac=2$ across various flow conditions. 
\begin{table}[b!!]
	\centering
	\begin{tabular}{|l|l|l|l|l|}
		\hline
		& $\mfac$ = 0.25 & $\mfac$ = 0.1 & $\mfac$ = 0.05& $\mfac$ = 0.05 -- 0.27   \\ \hline
		 $\CD $ \QS{Discrepancy}    & $1.15\times10^{-2}$   & $1.19\times10^{-2}$  & $1.19\times10^{-2}$&$1.16\times10^{-2}$\\ \hline
		$\CL$  \QS{Discrepancy}  & $2.51\times10^{-1}$   & $2.48\times10^{-1}$  & $2.48\times10^{-1}$&$2.51\times10^{-1}$\\ \hline
	\end{tabular}
	\caption{Averaged relative discrepancy of the $\CD$  and averaged RPD of the $\CL$ between the unmodified IIM~\cite{KOLAHDOUZ2020108854} with $\mfac = 2$ and the stabilized IIM with $\epsilon=116.5$ for smaller values of $\mfac$ over the time range $1$ to $100$.}
	\label{table: MSE_vary_Mfac}
\end{table}

\begin{table}[b!!]
	\centering
	\begin{tabular}{|l|l|l|l|}
		\hline
		& $\epsilon$ = 116.5 & $\epsilon$ = 11.65  & $\epsilon$ = 1.165    \\ \hline
		
		$\CD$ \QS{Discrepancy}  & $1.26\times10^{-2}$   & $3.14\times10^{-2}$  & $1.07\times10^{-1}$\\ \hline
		$\CL$ \QS{Discrepancy}  & $1.97\times10^{-1}$   & $3.56\times10^{-1}$  & $4.97\times10^{-1}$\\ \hline
	\end{tabular}
	\caption{Averaged relative discrepancy of the $\CD$ and averaged RPD of the $\CL$ relative percent difference between the unmodified IIM~\cite{KOLAHDOUZ2020108854} with $\mfac = 2$ and the stabilized IIM with different values of $\epsilon$ for $\mfac=0.05$ over the time range $1$ to $50$. }
	\label{table: MSE_vary_epsilon}
\end{table}
Figs.~\ref{fig:vary_eps} and \ref{fig:vary_eps2} and Table \ref{table: MSE_vary_epsilon} explore the effect of the choice of the stability coefficient $\epsilon$ in stabilized IIM. We observe that the oscillations in the simulation with the stabilized IIM decrease as $\epsilon$ increases within an appropriate range; however, choosing $\epsilon$ to be too large can also negatively impact both accuracy and stability.
 
\begin{figure}[!!b!!]
	\centering
	\hskip -.3cm
	\includegraphics[width=0.97\textwidth]{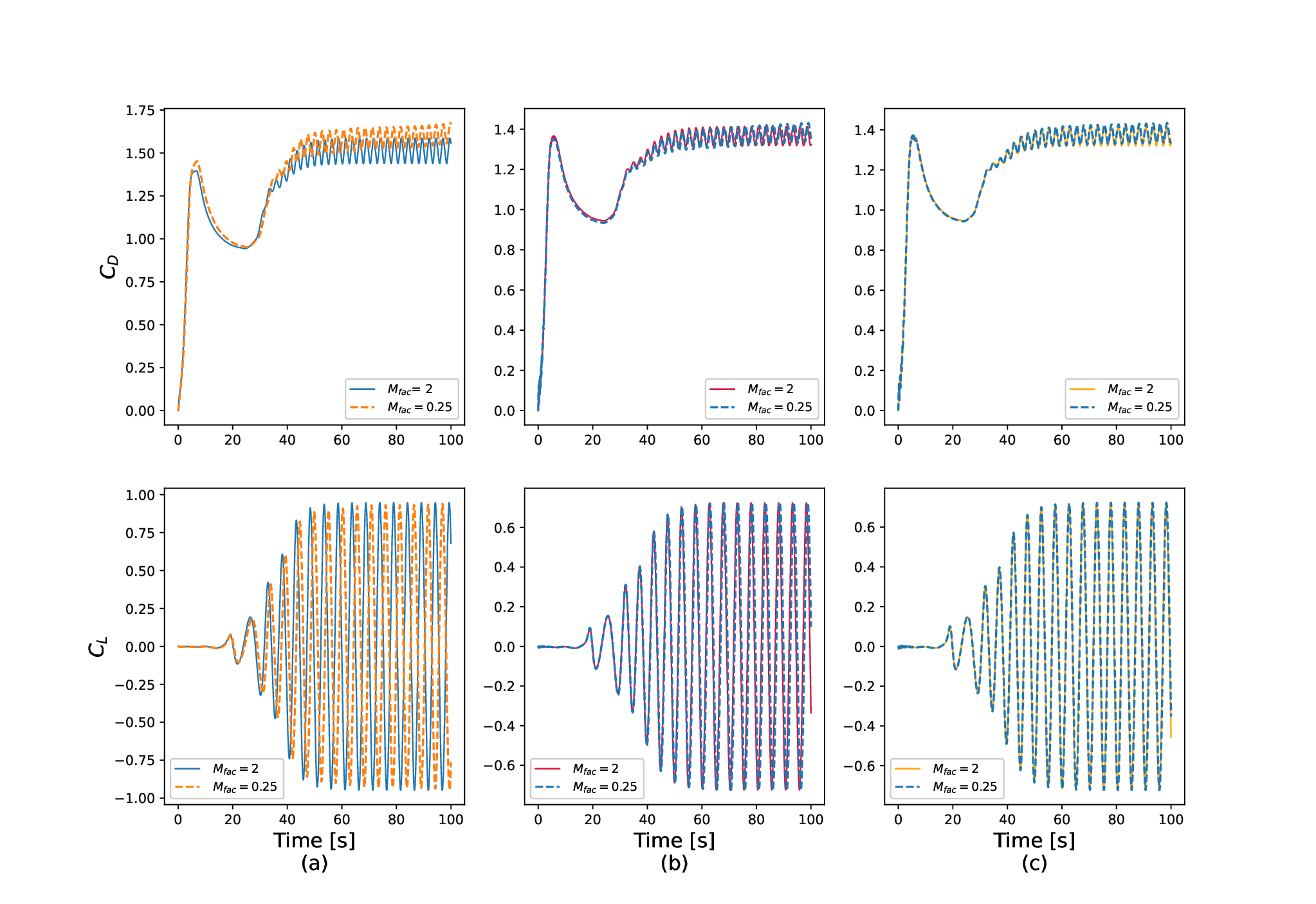}\quad
	\caption{{\color{black}Drag coefficient $\CD$ and lift coefficients $\CL$ over time  for two-dimensional flow past a cylinder with $\mfac=0.25$, (a) $\  h_{\text{finest}}=\frac{L}{1024}$,(b) $h_{\text{finest}}=\frac{L}{2048}$ and (c) $h_{\text{finest}}=\frac{L}{4096}$, generated by the stabilized IIM with $\epsilon=116.5$.}}
	\label{fig:2D_flow_cylinder_different_h}
\end{figure}

{\color{black}Results reported in Figs.~\ref{fig:vary_eps} and \ref{fig:vary_eps2} and Table \ref{table: MSE_vary_epsilon} can help to determine an appropriate value the parameter $\epsilon$. If the fluid solver diverges during the simulation and the interface experiences large spurious motions, usually it means the $\epsilon$ is not large enough to stablize the discrete system. However, if one noticed the dynamics of the simulations is changed, usually that indicates $\epsilon$ needs to be reduced.} {\color{black}Fig.~\ref{fig:2D_flow_cylinder_different_h} tests the scaling of the stability coefficient in Eq.~(\ref{tikhnov_eqn}). We observe that once a proper $\epsilon$ is found, the stabilized IIM reproduces nearly identical lift and drag coefficient dynamics for different values of the fluid mesh size $h$ as the unmodified IIM that satisfies $\mfac = 2$.}


\subsection{Flow past a stationary sphere}

\begin{figure}[b!!]
	\centering
	\hskip -.3cm
	\includegraphics[width=.36\textwidth]{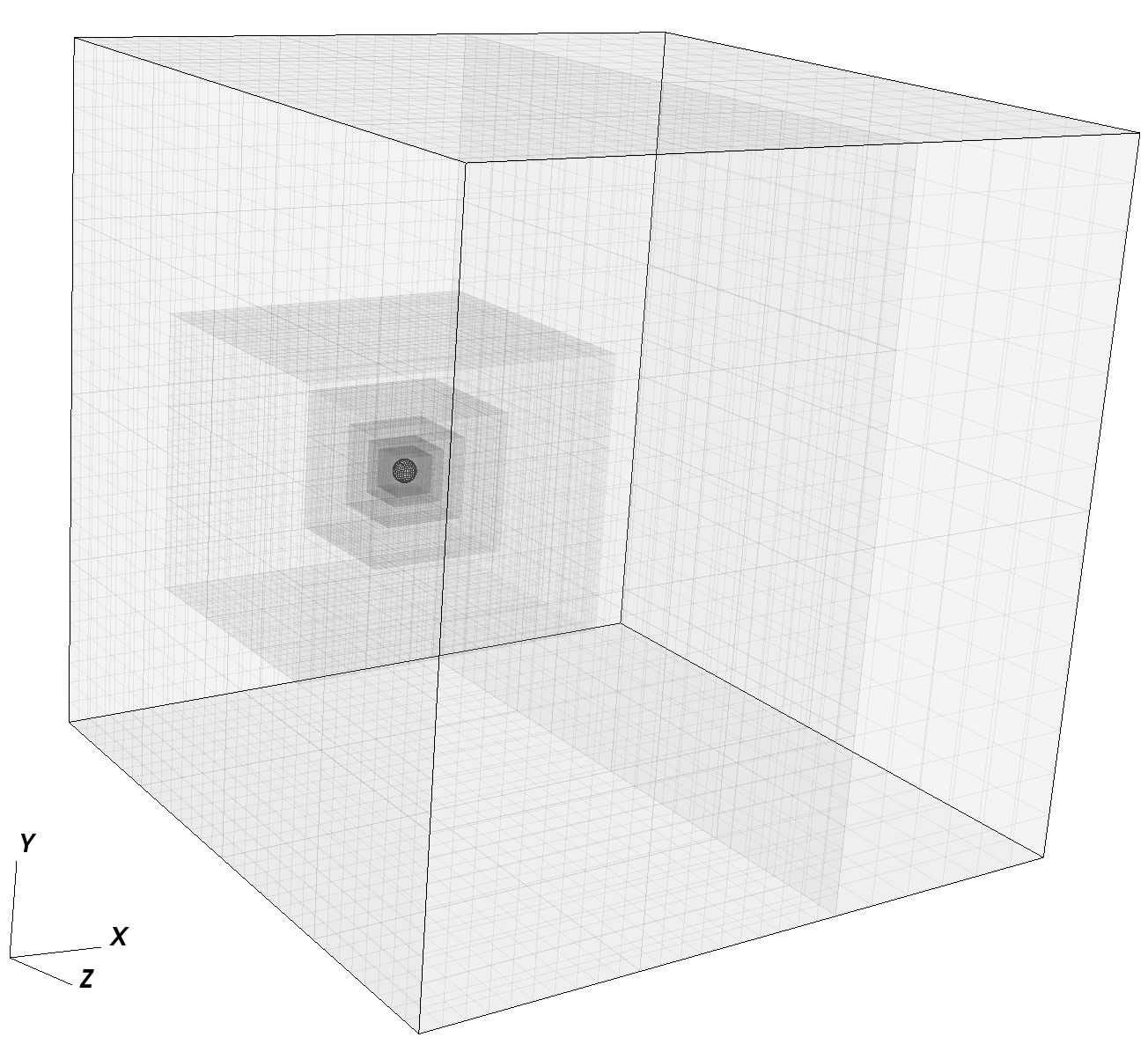}
	\includegraphics[width=.36\textwidth]{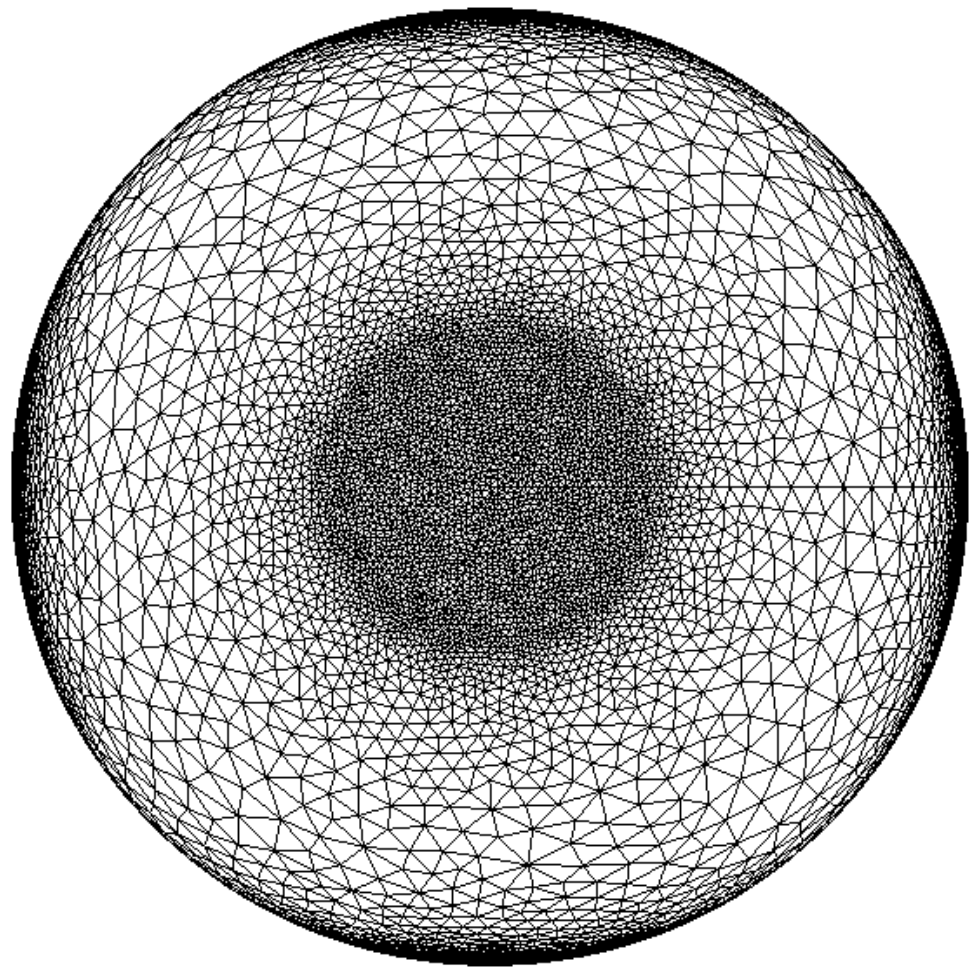}
	\caption{Computational meshes used to simulate flow past a stationary sphere. Fluid mesh (left) and interface mesh (right), which is designed with a broad range of element sizes.}
	\label{fig:mesh_sphere}
\end{figure}
This test investigates robustness of the stabilized IIM in a 3D setup. The computational domain is $\Omega=[-15, 45] \times [-30, 30] \times [-30, 30]$, which is a cube with length $L=60$. The immersed boundary is a sphere centered at the origin with a diameter of $1$. We impose inflow velocity boundary condition $\bu=(\frac{\tanh(t/10-2)+\tanh(2)}{1+\tanh(2)},0,0)$ on the boundary $(x = -15)$ , zero normal traction and zero tangential velocity is imposed at the right
boundary $(x = 45)$ as an outflow condition.  Along the bottom $(y = -30)$,  top $(y = 30)$, front $(z = 30)$, and back $(z = -30)$ boundaries, the
normal velocity and tangential traction are set to zero. 
We choose $\rho = 1$ and use inflow velocity $(1,0,0)$ as the characteristic velocity. The Reynolds number is $\Re=\frac{\rho U D}{\mu}$, $\mu=\frac{1}{Re}$. Reynolds numbers of 20, 100, and 200 are considered. 

The computational domain is discretized using 7 levels of local adaptive refinement, with a refinement ratio of $r = 2$.
The coarsest Cartesian grid spacing is set as $h_{\text{coarsest}}=\frac{L}{16}$ and the finest is  $h_{\text{finest}}=\frac{L}{2048}$; see Fig.~\ref{fig:mesh_sphere} (left). We {\color{black}consider} a non-uniform mesh with a range of element sizes, so that $\mfac$ is in the range $0.1$ -- $1$; see Fig.~\ref{fig:mesh_sphere} (right). The time size is set to $\Delta t = 0.00125$. To assess the dynamics from the numerical simulations, we compute nondimensional quantities including the drag coefficient $\CD$ and lift coefficient $\CL^y, \CL^z$ defined as,
\begin{align}
	(\CD,\CL^y,\CL^z)=\frac{-2\int_{\Gamma_0}\bF(\bX,t)\, \mathrm{d}A}{A_{\text{proj}}},
\end{align}
in which $A_{\text{proj}}=\frac{\pi}{4}$ is the projected area of the sphere with diameter $D = 1$. Without stabilization, the unmodified IIM~\cite{KOLAHDOUZ2020108854} exhibits severe instabilities caused by locally small values of $\mfac$, similar to what was demonstrated in the previous section. Table \ref{tb:3D_flow_cylinder} compares the results of the unmodified IIM, which satisfies the $\mfac > 1$ constraint with $\mfac = 2$, and the stabilized IIM for the given nonuniform surface mesh to previous work~\cite{XU20082068, Fornberg_1988, TURTON198683, FADLUN200035, Campregher09} for Reynolds numbers ranging from 20 to 200. We observe good agreement between the results from the stabilized IIM with $\mfac = 0.1$--$1$ and the reported results from previous studies. 
\begin{table}[t]
	\centering
	\begin{tabular}{|l|l|l|l|}
		\hline
		& \Re = 20 & \Re = 100 & \Re = 200 \\ \hline
		Xu and Wang~\cite{XU20082068}& 2.73   & 1.15  & 0.88\\ \hline
		Fornberg.~\cite{Fornberg_1988}& -   & 1.0852 & 0.7683 \\ \hline
		Turton and Levenspiel~\cite{TURTON198683}& 2.6866   & 1.0994 & 0.8025  \\ \hline
		Fadlun et al.~\cite{FADLUN200035}& - &  1.0794 & 0.7567\\ \hline
		Campregher et al.~\cite{Campregher09}& - &1.1781&0.8150 \\ \hline
		IIM without stabilization~\cite{KOLAHDOUZ2020108854} $\mfac=2$&2.7232&1.07966&0.7566 \\ \hline
		Stabilized IIM with $\epsilon=102.4$ $\mfac=0.1-1$&2.7236&1.0711&0.7447 \\ \hline
	\end{tabular}
	\caption{Drag coefficients for three-dimensional flow past a sphere at various Reynolds numbers are simulated using the stabilized IIM and compared with previous computational work~\cite{XU20082068, Fornberg_1988, TURTON198683, FADLUN200035}, as well as empirical data~\cite{Campregher09}. }
	\label{tb:3D_flow_cylinder}
\end{table}



\subsection{Vortex-induced vibration of a cylinder}

Next we investigate the robustness of the stabilized IIM in a more complex FSI model setup. In this section, the motion $\bxi(\bX, t)$ of the material interface $\Gamma_0$ is governed by the equations of rigid-body motion. For more technical details regarding the FSI coupling scheme, we refer to our previous work~\cite{KOLAHDOUZ2021110442}. The problem of viscous flow past an elastically mounted two-dimensional cylinder undergoing vortex-{\color{black}induced} vibration
(VIV) has been extensively studied through both numerical simulations and experimental investigations. This is attributed to its wide-ranging engineering applications and the complex vortex dynamics involved. Furthermore, this problem serves as a significant benchmark for evaluating fluid-structure interaction (FSI) algorithms~\cite{AHN2006671,Borazjani2008,Bao2012, Jianming2008,Jianming2015,KIM2018296}. The governing equations for the motion of the cylinder with two degrees of freedom are:
\begin{align}
	M_\text{s}\Ddot{d}_\text{c}^x+C_\text{s} \dot{d}_\text{c}^x+K_\text{s} d_\text{c}^x&=f^x,\\
	M_\text{s}\Ddot{d}_\text{c}^y+C_\text{s} \dot{d}_\text{c}^y+K_\text{s} d_\text{c}^y&=f^y,
\end{align}
in which $d^x_\text{c}$ and $d^y_\text{c}$ are the horizontal and vertical displacements of the cylinder's center of mass, respectively. The mass per unit length of the cylinder is denoted by $M_\text{s}$, and $C_\text{s}$ and $K_\text{s}$ are the damping and stiffness constants of the spring. $f_x$ and $f_y$ are the instantaneous
drag and lift forces. To compare with previous work~\cite{KOLAHDOUZ2021110442}, we define the non-dimensional horizontal and vertical displacements of the cylinder's center in the streamwise and transverse directions as $\hat{d}^x_\text{c} = \frac{d^x_\text{c}}{D}$ and $\hat{d}^y_\text{c} = \frac{d^y_\text{c}}{D}$, in which $D$ is the diameter of the cylinder. We set $U_\infty$ to be the free stream flow velocity. The mass ratio and reduced velocity are $m^* = \frac{\rho_\text{s}}{\rho_\text{f}}$ and $U^* = \frac{U_{\infty}}{f_\text{n} D}$, in which $f_\text{n} = \frac{\sqrt{K_\text{s}/M_\text{s}}}{2\pi}$ is the natural frequency of the structure. The damping ratio is $\gamma = \frac{C_\text{s}}{2\sqrt{K_\text{s} M_\text{s}}}$.
We consider the benchmark problem of a circular cylinder undergoing VIV. We are interested in capturing the well-characterized vortex “lock-in” phenomenon observed in previous work~\cite{AHN2006671,Borazjani2008,Bao2012,Blackburn1993TwoAT}. Within the lock-in regime, the vortex shedding frequency closely matches the natural frequency of the structure, resulting in large amplitude vibrations. Physical parameters are chosen to match the benchmark results in Blackburn and Karniadakis~\cite{Blackburn1993TwoAT}. The computational domain is $\Omega = [-30\ \textrm{cm}, 45\ \textrm{cm}] \times [-30\ \textrm{cm}, 30\ \textrm{cm}]$. The cylinder has diameter $D = 1\ \textrm{cm}$, is initially at rest, and is centered at the origin. A uniform inflow velocity $U = (1\ \textrm{cm} \cdot s^{-1}, 0\ \textrm{cm} \cdot s^{-1})$ is imposed on the left boundary $(x = -30\ \textrm{cm})$,  and zero normal traction and tangential velocity outflow conditions are imposed at the right boundary $(x = 45\ \textrm{cm})$.
Along the bottom $(y = -30\ \textrm{cm})$ and top $(y = 30\ \textrm{cm})$  boundaries, zero normal velocity and tangential traction are imposed.

The computational domain is discretized using $N = 6$ nested grid levels, with coarse grid spacing
$h_{\textrm{coarsest}} =\frac{L_y}{128} = 0.46875\ \textrm{cm}$ and refinement ratio $r = 2$ between levels, leading to $h_{\text{finest}} = 0.0145\ \textrm{cm}$.  The Reynolds number $\Re = \frac{\rho fU_{\infty}D}{\mu f}$ is fixed at 200, the damping is set to zero $(\gamma = 0)$, and the mass ratio is $m^* = \frac{4}{\pi}$.

\begin{figure}[b!!]
	\centering
	\hskip -.3cm
	\includegraphics[width=.80\textwidth]{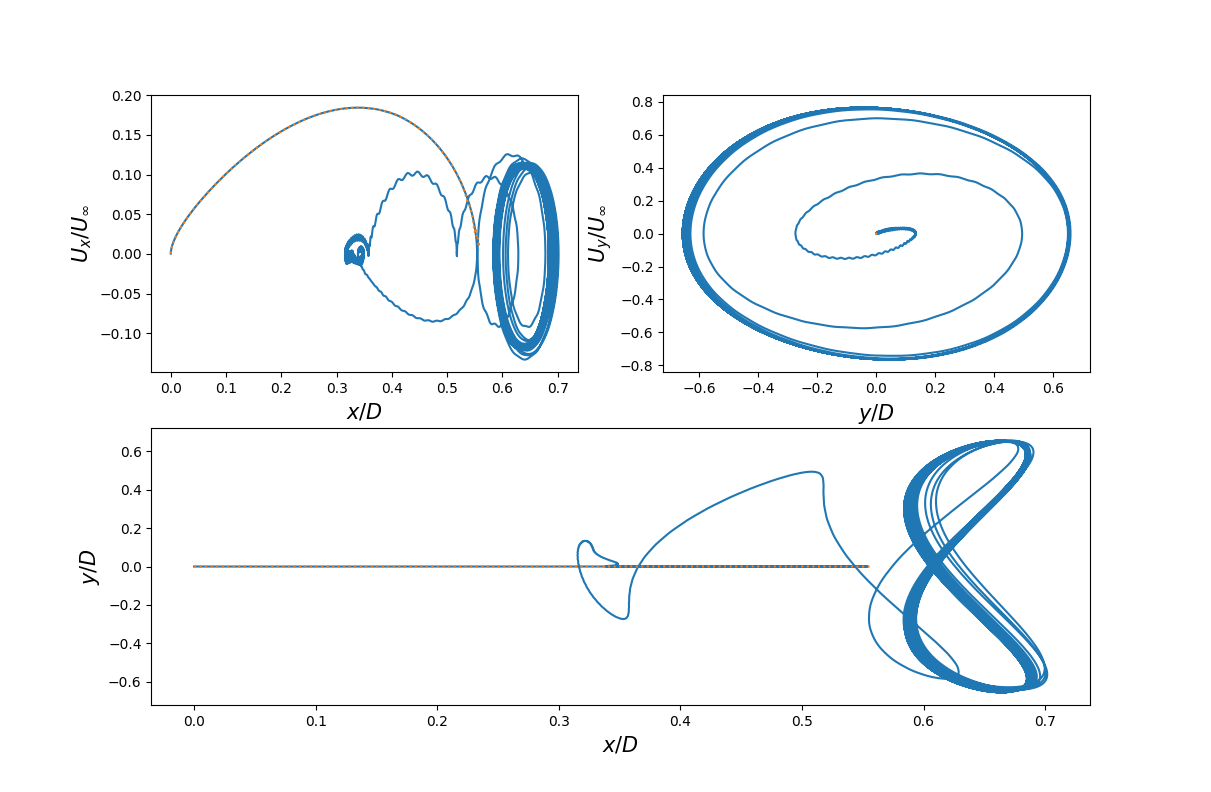}
	\caption{Phase plots of the center of mass displacement and velocity responses for an elastically mounted
		cylinder with mass ratio of $m^* = 4 / \pi$ obtained by IIM without stabilization~\cite{KOLAHDOUZ2021110442}. Other simulation parameters include $U^* = 5, \gamma = 0.01$, and $\Re = 200$. Solid line: $\mfac=2$; dashed line: $\mfac=0.1$. {\color{black}Simulation with $\mfac = 0.1$ stopped because of excessive spurious interfacial motions.}}
	\label{fig:no_s_ex4_4}
\end{figure}
\begin{figure}[t!!]
	\centering
	\hskip -.3cm
	\includegraphics[width=.70\textwidth]{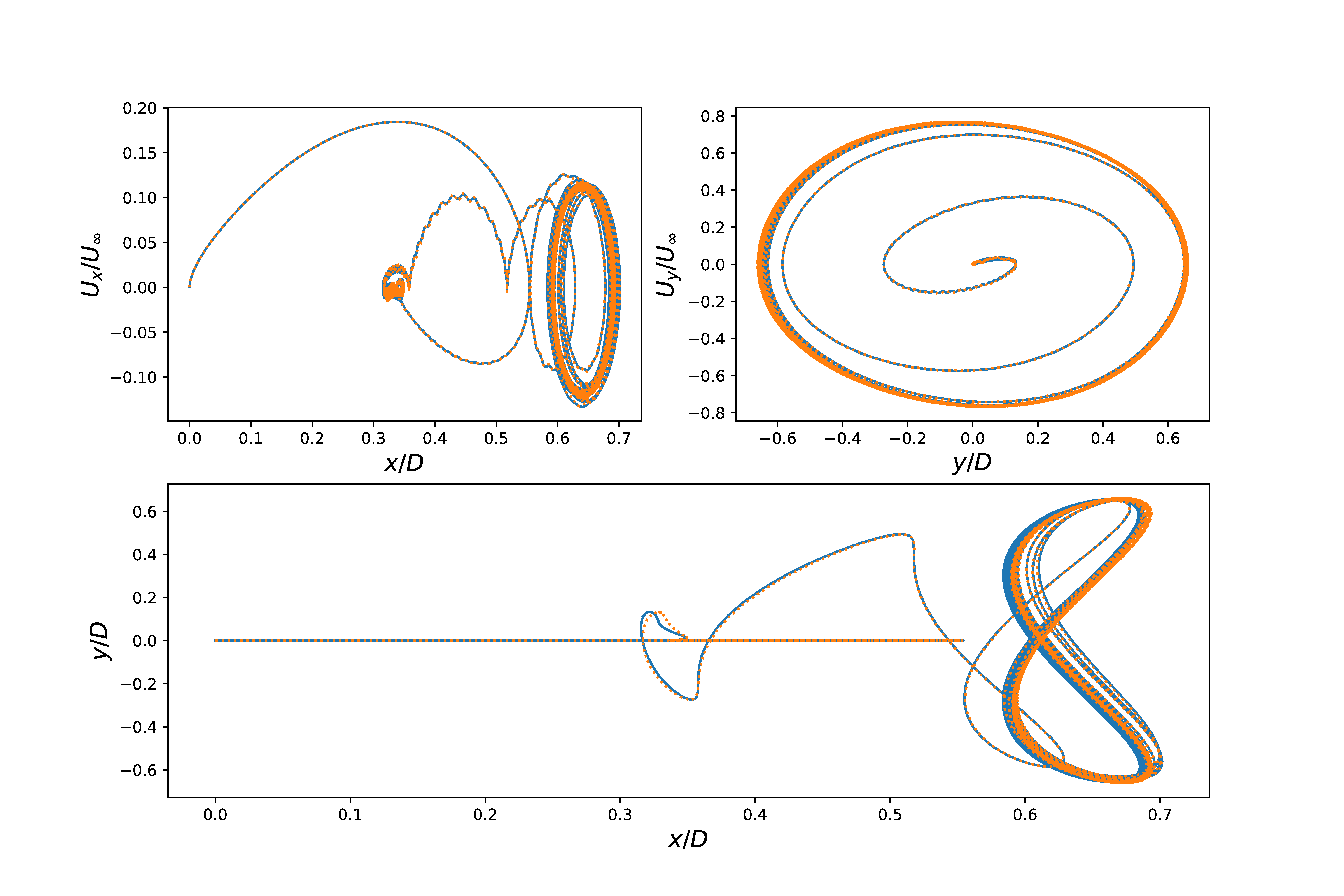}
	\caption{Phase plots of the center of mass displacement and velocity responses for an elastically mounted
		cylinder with mass ratio of $m^* = 4 / \pi$ . Other simulation parameters include $U^* = 5, \gamma = 0.01$, and $\Re = 200$. Solid line: IIM without stabilization~\cite{KOLAHDOUZ2021110442} with $\mfac=2$; dashed line:  stabilized-IIM with $\epsilon=46.6 $, $\mfac=0.1$. The figure shows that our stabilized IIM is robust with respect to the mesh size ratio $\mfac=0.1$, reproducing dynamics nearly identical to the unmodified IIM with $\mfac=2$.}
	\label{fig:ex4_4}
\end{figure}
\begin{figure}[b!!]
	\centering
	\hskip -.3cm
	\includegraphics[width=.70\textwidth]{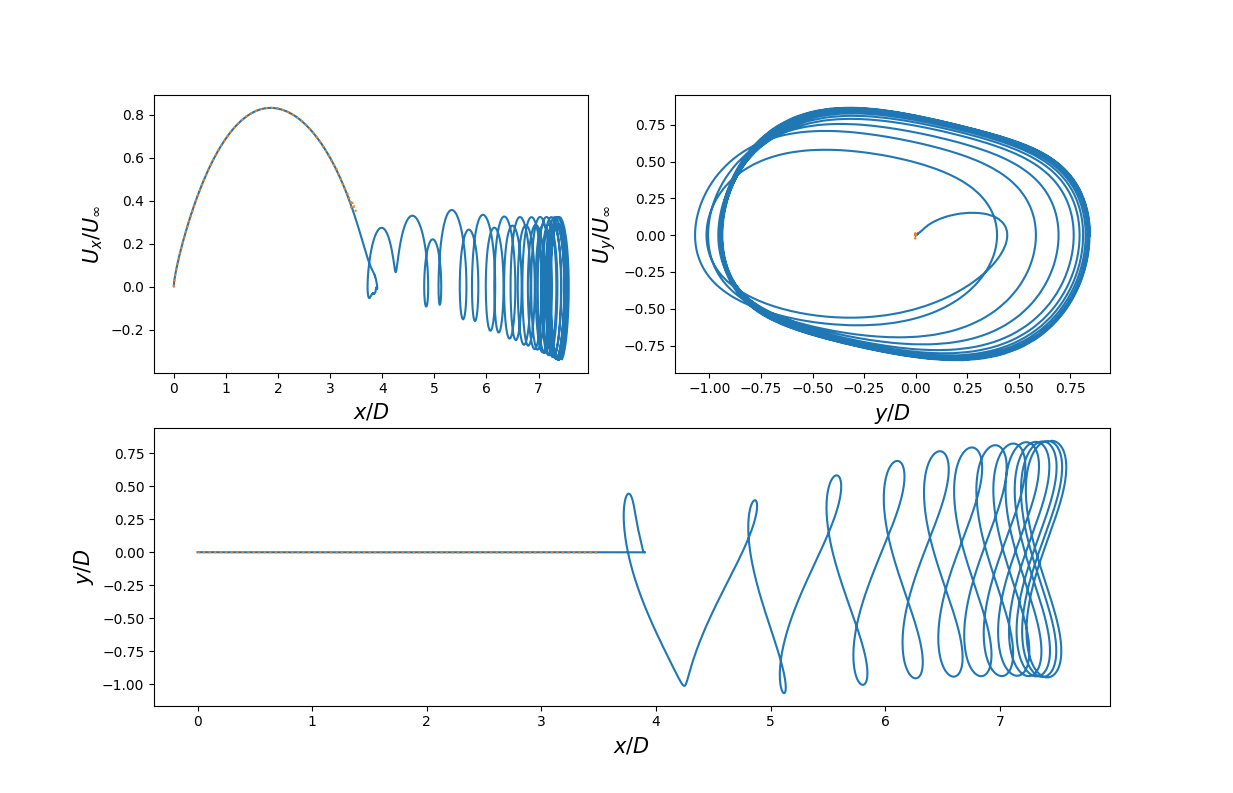}
	\caption{Phase plots of the center of mass displacement and velocity responses for an elastically mounted
		cylinder with mass ratio of $m^* = 0.4 / \pi$ obtained by IIM without stabilization~\cite{KOLAHDOUZ2021110442}. Other simulation parameters include $U^* = 5$,
		$\gamma = 0.01$, and $\Re = 200$. Solid line: $\mfac=2$; Dashed line: $\mfac=0.1$. {\color{black}Simulation with $\mfac = 0.1$ stopped because of excessive spurious interfacial motions.}}
	\label{fig:no_s_ex4_04}
\end{figure}
\begin{figure}[t!!]
	\centering
	\hskip -.3cm
	\includegraphics[width=.70\textwidth]{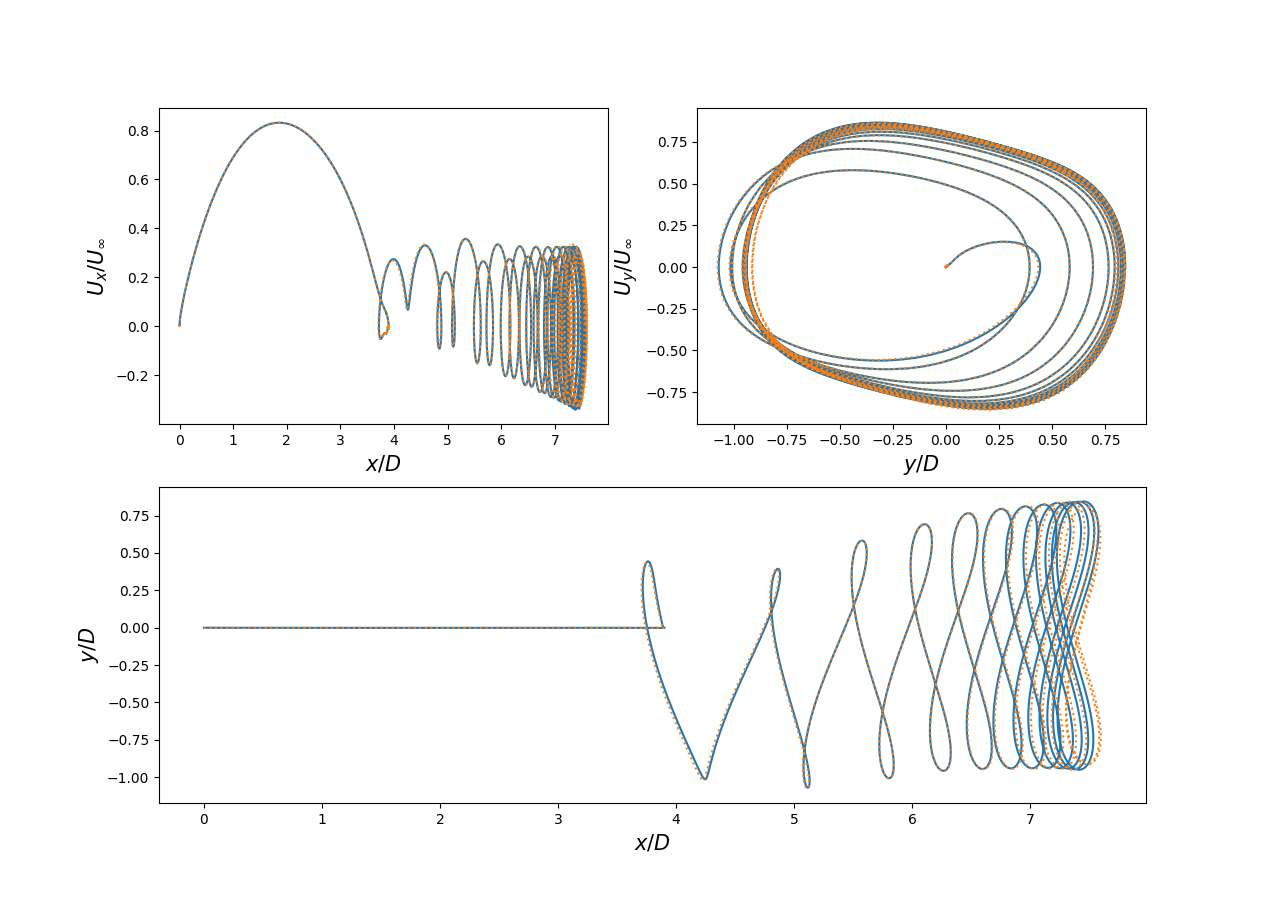}
	\caption{Phase plots of the center of mass displacement and velocity responses for an elastically mounted
		cylinder with mass ratio of $m^* = 0.4 / \pi$. Other simulation parameters include $U^* = 5$,
		$\gamma = 0.01$, and $\Re = 200$. Solid line: IIM without stabilization~\cite{KOLAHDOUZ2021110442} with $\mfac=2$; dashed line:  stabilized-IIM with $\epsilon=46.6 $, $\mfac=0.1$. The figure shows that our stabilized IIM is robust with respect to the mesh size ratio $\mfac=0.1$, reproducing dynamics nearly identical to the unmodified IIM with $\mfac=2$.}
	\label{fig:ex4_04}
\end{figure}
The vortex shedding generated by the oscillating cylinder is expected to produce a periodic `figure-eight' pattern. Fig.~\ref{fig:no_s_ex4_4} shows centerline trajectory ($x/D$-$y/D$) and the dimensionless displacement velocity phases ($x/D$-$U_x/U_{\infty}$ and $y/D$-$U_y/U_{\infty}$) for both the unmodified IIM with $\mfac = 2$ and $\mfac = 0.1$. We observe the unmodified IIM simulation with $\mfac=0.1$ soon exhibits severe instabilities. In contrast, the stabilized IIM with $\mfac = 0.1$, as shown in Fig.~\ref{fig:ex4_4}, successfully generates dynamics comparable to the unmodified IIM with $\mfac=2$. The phase response obtained from our simulations shows good alignment with each other. We compute the averaged RPD to quantify the discrepancy between the unmodified IIM with $\mfac = 2$ and the stabilized IIM with $\mfac = 0.1$.  The averaged RPD of the center of mass displacement is $1.44$, and the averaged RPD of the center of mass velocity is $3.16$. The results from the stabilized IIM with $\mfac = 0.1$ and the unmodified IIM with $\mfac = 2$ show good agreement with previous observations reported by Yang and Stern~\cite{Jianming2015}, Blackburn and Karniadakis~\cite{Blackburn1993TwoAT}, and Liu and Hu~\cite{Liuhu18}.

We next investigate lower mass ratios compared to the one examined earlier. Specifically, we consider a case with a density ratio of $m^*=0.4/\pi$. The remaining simulation parameters are the same as in the previous example settings. Fig.~\ref{fig:no_s_ex4_04}
shows the centerline trajectory ($x/D$-$y/D$) and the dimensionless displacement velocity phases ($x/D$-$U_x/U_{\infty}$ and $y/D$-$U_y/U_{\infty}$) for both the unmodified IIM with $\mfac = 2$ and $\mfac = 0.1$. The unmodified IIM with $\mfac = 0.1$ exhibits severe instabilities soon after the beginning, while the stabilized IIM with $\mfac = 0.1$, shown in Fig.~\ref{fig:ex4_04}, successfully replicates dynamics comparable to the unmodified IIM with $\mfac = 2$. The phase responses from both simulations align well. To quantify the discrepancy, we compute the averaged RPD between the unmodified IIM with $\mfac = 2$ and the stabilized IIM with $\mfac = 0.1$. The averaged RPD for the center of mass displacement is $0.83$, and for the center of mass velocity, it is $1.84$. Figs.~\ref{fig:no_s_ex4_04} and \ref{fig:ex4_04} shows that our stabilized IIM is robust with respect to the mesh size ratio $\mfac=0.1$ and reproduces nearly identical dynamics compared to IIM without stabilization~\cite{KOLAHDOUZ2021110442} with $\mfac=2$.

\subsection{Two-dimensional cylinder in shear flow}

\begin{figure}[b!!]
	\centering
	\hskip -.3cm
	\includegraphics[width=.4\textwidth]{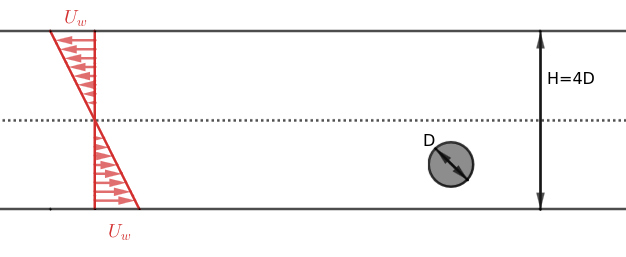}
	\caption{Two-dimensional cylinder in shear flow problem setup.}
	\label{fig:cylinder_shear_flow}
\end{figure}
\begin{figure}[t!!]
	\centering
	\hskip -.3cm
	\includegraphics[width=.70\textwidth]{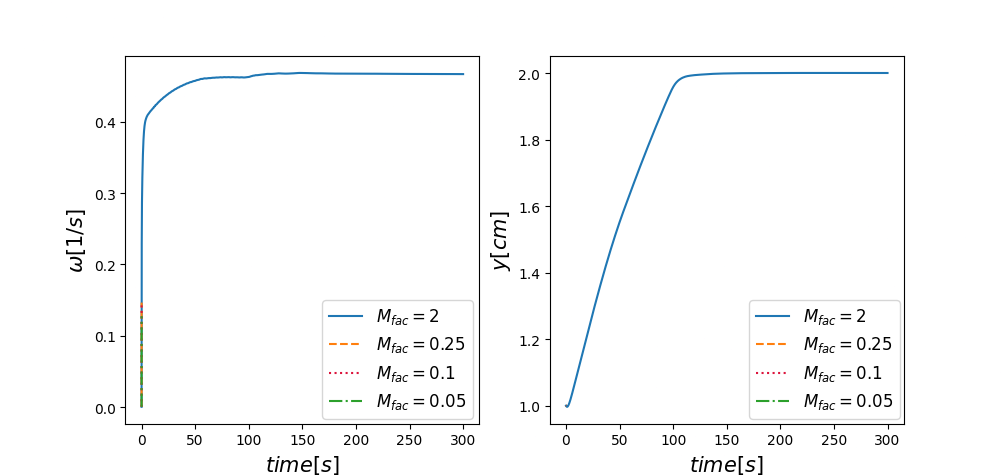}
	\caption{Two-dimensional cylinder in shear flow results
		obtained using IIM without stabilization~\cite{KOLAHDOUZ2020108854} with $\mfac=2,0.25,0.1,$ and $0.05$. (Left) Time history of the angular velocity of the cylinder. The angular velocity converges to
		$\omega_r = 0.464 s^{-1}$. (Right) Time history of the
		$y$ position of the cylinder center of mass. {\color{black}Simulations with $\mfac = 0.25,0.1,$ and $0.05$ stopped because of excessive spurious interfacial motions.}}
	\label{fig:cylinder_shear_flow_original}
\end{figure}
\begin{figure}[hb!!]
	\centering
	\hskip -.3cm
	\includegraphics[width=.70\textwidth]{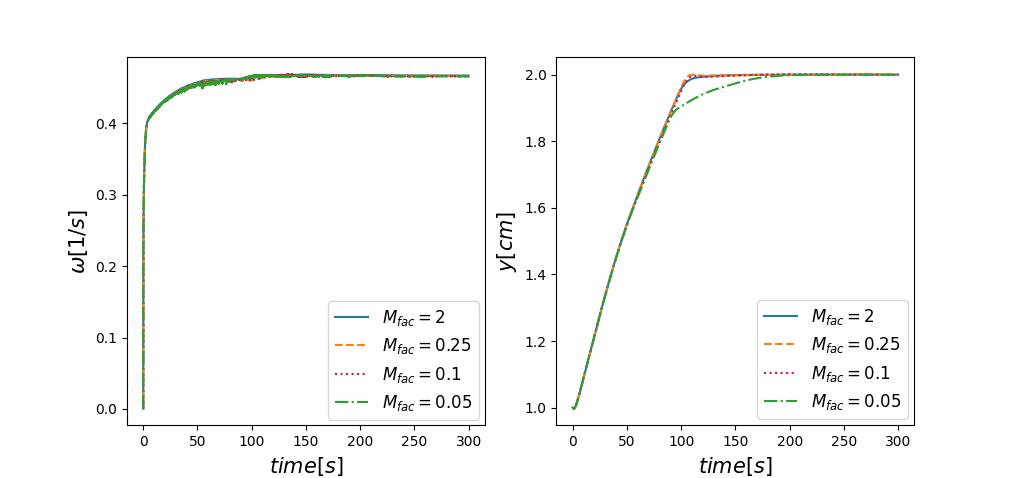}
	\caption{Two-dimensional cylinder in shear flow results
		obtained using unmodified IIM with $\mfac=2$ and stabilized IIM with $\epsilon=1.024$, $\mfac=0.25,0.1,$  and $0.05$.  (Left) Time history of the angular velocity of the cylinder. The angular velocity converges to
		$\omega_r = 0.464 s^{-1}$. (Right) Time history of the
		$y$ position of the cylinder center of mass. The figure shows that our stabilized IIM is robust with respect to the mesh size ratio $\mfac=0.25, 0.1,$ and $0.05$, reproducing dynamics nearly identical to the unmodified IIM with $\mfac=2$.}
	\label{fig:cylinder_shear_flow_stablized}
\end{figure}
\begin{table}[ht!!]
	\centering
	\begin{tabular}{|l|l|l|l|}
		\hline
		& $\mfac$ = 0.25 & $\mfac$ = 0.1 & $\mfac$ = 0.05 \\ \hline
		$\omega$ \QS{Discrepancy}  & $2.99\times10^{-3}$   & $3.58\times10^{-3}$  & $4.06\times10^{-3}$\\ \hline
		$y$ \QS{Discrepancy}  & $6.68\times10^{-4}$   & $1.14\times10^{-3}$  & $6.26\times10^{-3}$\\ \hline
	\end{tabular}
	\caption{Averaged RPD between the unmodified IIM~\cite{KOLAHDOUZ2020108854} with $\mfac = 2$ and the stabilized IIM with $\epsilon=1.024$ for $\mfac=0.25,0.1,$ and $0.05$. }
	\label{fig:2D_shear_ball}
\end{table}

This numerical test considers a circular cylinder in shear flow, as studied by Feng
et al.~\cite{Feng_Hu_Joseph_1994} and Koladouz et al.~\cite{KOLAHDOUZ2020108854}.  The computational domain is defined as $\Omega = [-\textrm{L}/2, \textrm{L}/2] \times [0, \textrm{H}]$. The immersed interface is a cylinder that has a diameter of $\textrm{D} = 1\ \textrm{cm}$ , with the channel dimensions being $H = 4D$ in height and $L = 160D$ in length. A periodic boundary condition was applied in the $x$-direction, while the top and bottom walls move in the $x$ direction with velocities of $-U_w$ and $U_w$, respectively, resulting in a constant shear rate of $\dot{\gamma} = \frac{2U_w}{H}$. The Reynolds number is $\Re = \frac{\rho U_w L}{ \mu} = 40$. Initially, the cylinder is placed at $(x, y) = (0, \frac{H}{4})$ and released with zero initial velocity and rotation; see Fig.~\ref{fig:cylinder_shear_flow}. 

The computational domain is discretized using $N = 4$ nested grid levels, in which the coarsest grid spacing is $h_{\text{coarsest}} = \frac{H}{16}$, with a refinement ratio of $r=2$ between grid levels, $h_\text{finest} = \frac{\textrm{H}}{128}$.  The spring constant is $\kappa = 14000$, and the time step size is $\Delta t = 0.001$. 

The cylinder begins to rotate and gradually migrates toward the center of the channel. As shown in Figs.~\ref{fig:cylinder_shear_flow_original} and  \ref{fig:cylinder_shear_flow_stablized}, the stabilization is necessary for the IIM to maintain robustness with respect to the mesh size ratios $\mfac = 0.25, 0.2,$ and $ 0.05$. With these relative mesh spacing, our stabilized IIM produces dynamics nearly identical to the unmodified IIM approach~\cite{KOLAHDOUZ2021110442} for $\mfac>1$. To quantify the difference between the angular velocity of the cylinder $\Omega$ and the position of the cylinder from the unmodified IIM method with $\mfac = 2$ and the stabilized IIM with smaller values of $\mfac$, we employ averaged RPD to evaluate the discrepancy, see Table \ref{fig:2D_shear_ball}. We see a very small discrepancy between unmodified IIM with $\mfac = 2$ and the stabilized IIM with smaller values of $\mfac$. The angular velocity $\omega_r$ has been previously reported to be $0.47$~\cite{LACIS2016300,Feng_Hu_Joseph_1994,BHALLA2013446}. As demonstrated in Fig.~\ref{fig:cylinder_shear_flow_stablized}, our simulation captures the constant angular velocity, with a steady-state value of $\omega_r = 0.467-0.468$ across different values of $\mfac$, which is in excellent agreement with previous work~\cite{LACIS2016300,Feng_Hu_Joseph_1994,BHALLA2013446}.

{\color{black}\subsection{Soft disk in a lid driven cavity}
	\begin{figure}[hbt!]
		\centering
		\hspace{-0.3cm}
		
		\begin{subfigure}[b]{0.04\textwidth}
			\includegraphics[width=\textwidth]{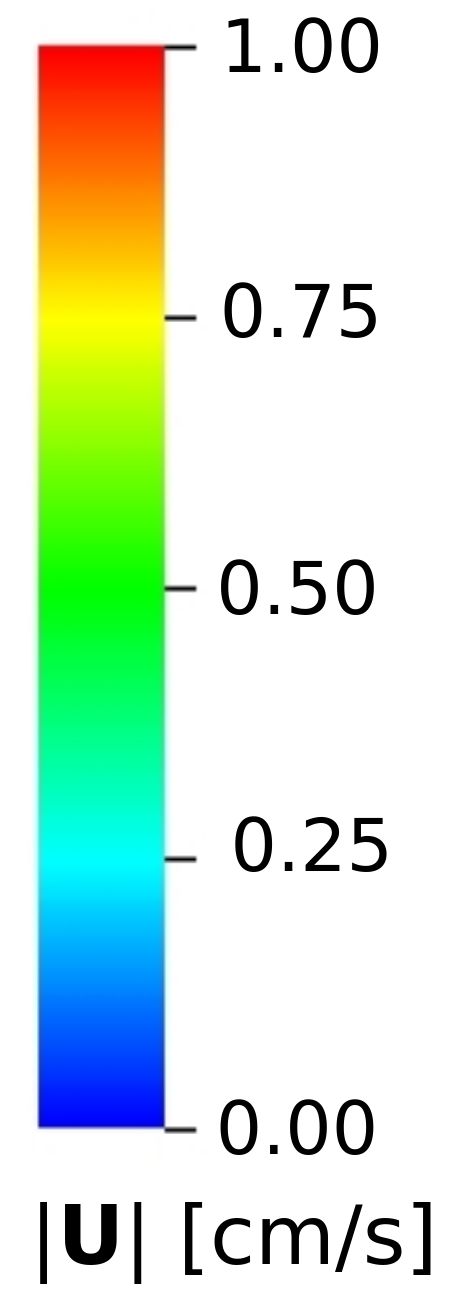}
			\caption*{} 
		\end{subfigure}
		\begin{subfigure}[b]{0.18\textwidth}
			\includegraphics[width=\textwidth]{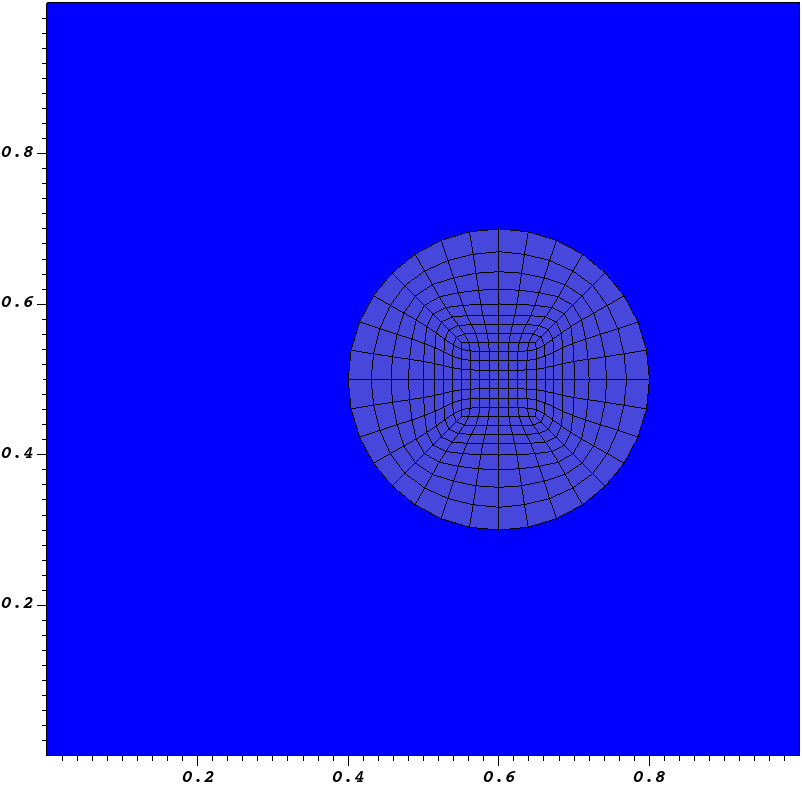}
			\caption{}
		\end{subfigure}
		\begin{subfigure}[b]{0.18\textwidth}
			\includegraphics[width=\textwidth]{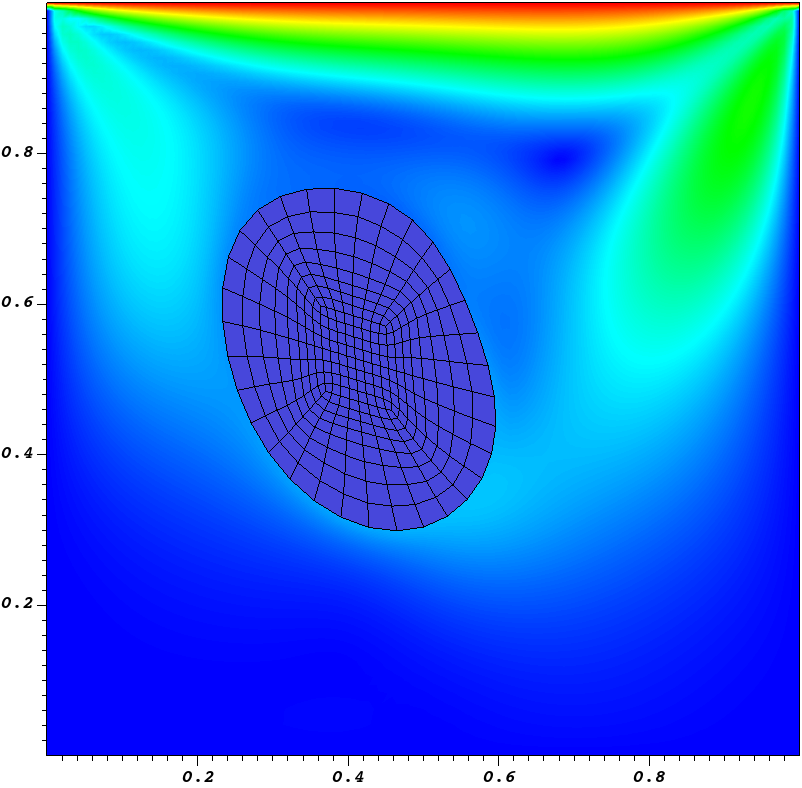}
			\caption{}
		\end{subfigure}
		\begin{subfigure}[b]{0.18\textwidth}
			\includegraphics[width=\textwidth]{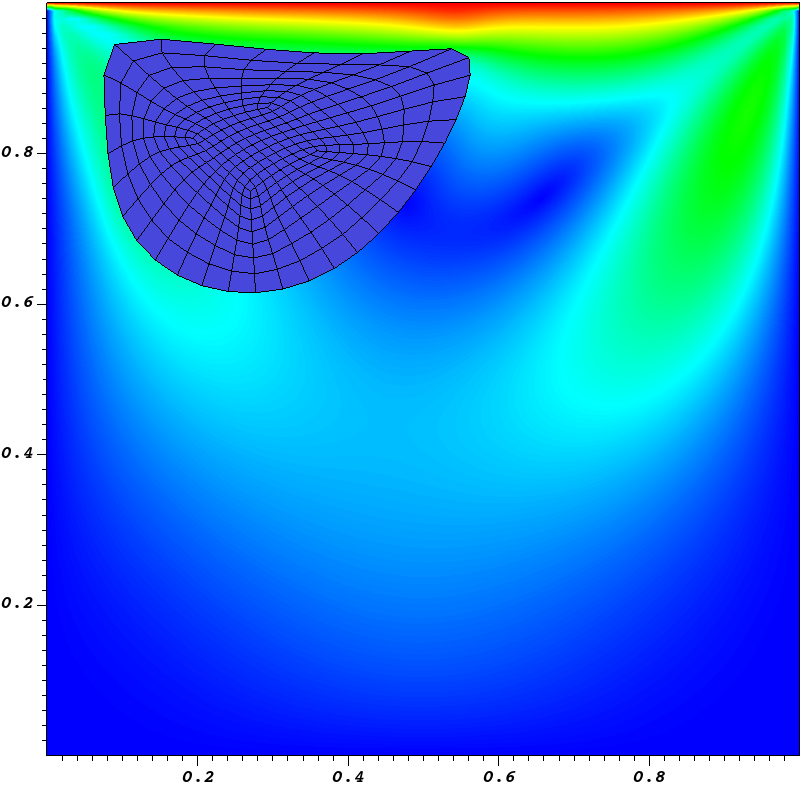}
			\caption{}
		\end{subfigure}
		\begin{subfigure}[b]{0.18\textwidth}
			\includegraphics[width=\textwidth]{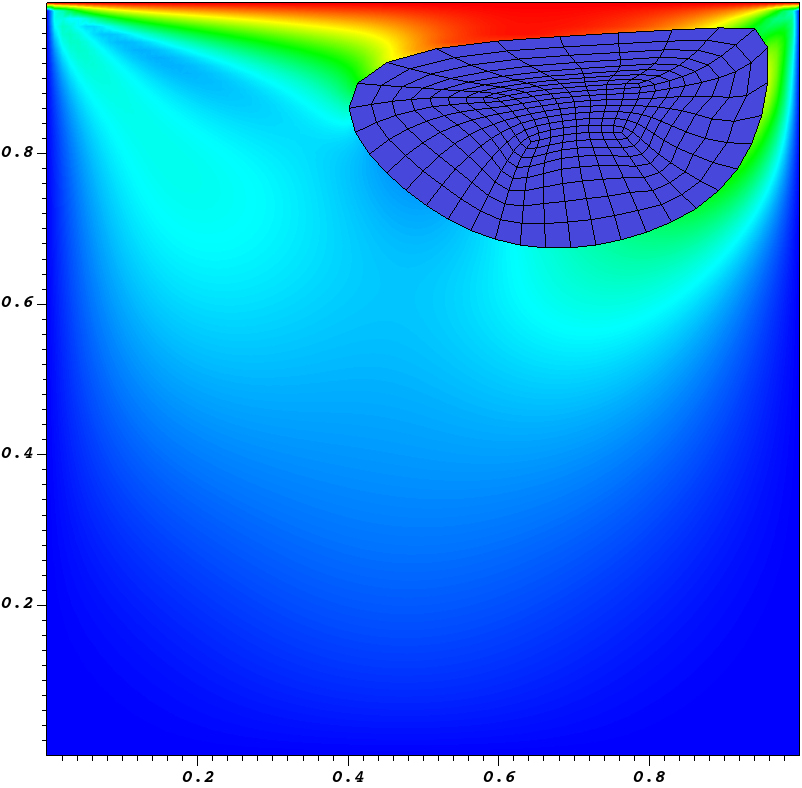}
			\caption{}
		\end{subfigure}
		\begin{subfigure}[b]{0.18\textwidth}
			\includegraphics[width=\textwidth]{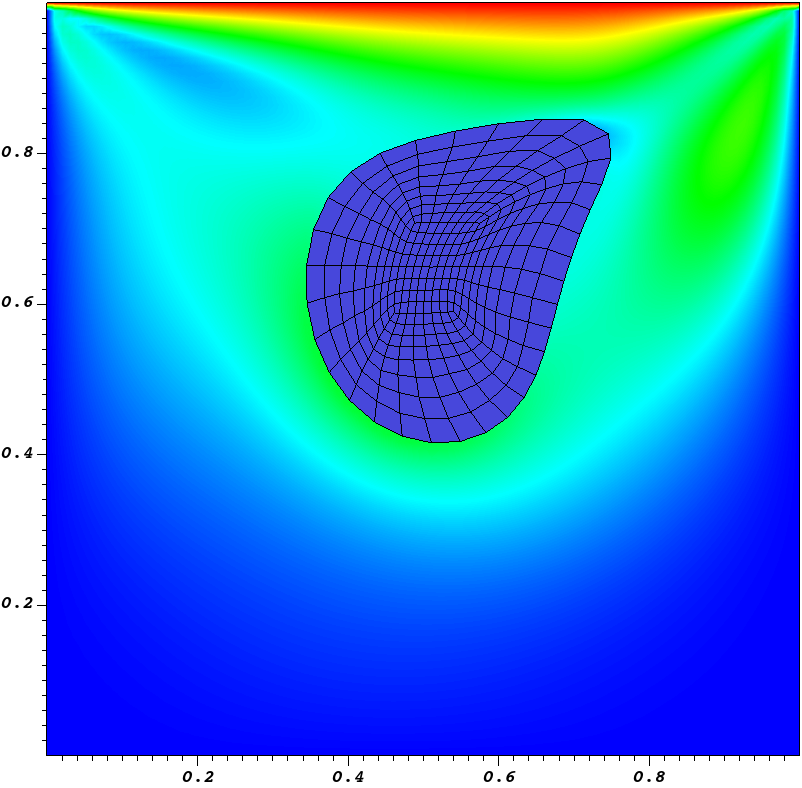}
			\caption{}
		\end{subfigure}
		
		\begin{subfigure}[b]{0.04\textwidth}
			\includegraphics[width=\textwidth]{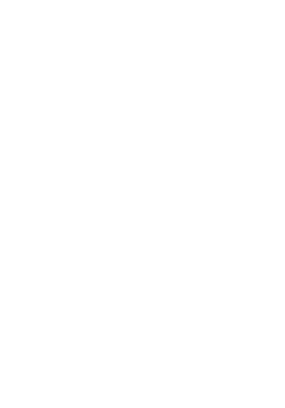}
			\caption*{}
		\end{subfigure}
		\begin{subfigure}[b]{0.18\textwidth}
			\includegraphics[width=\textwidth]{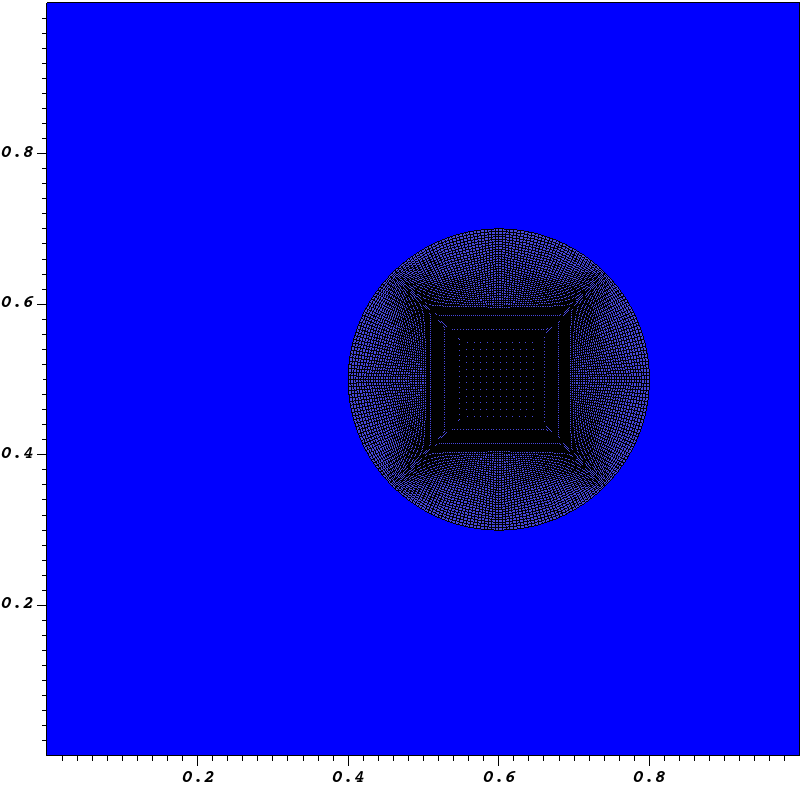}
			\caption{}
		\end{subfigure}
		\begin{subfigure}[b]{0.18\textwidth}
			\includegraphics[width=\textwidth]{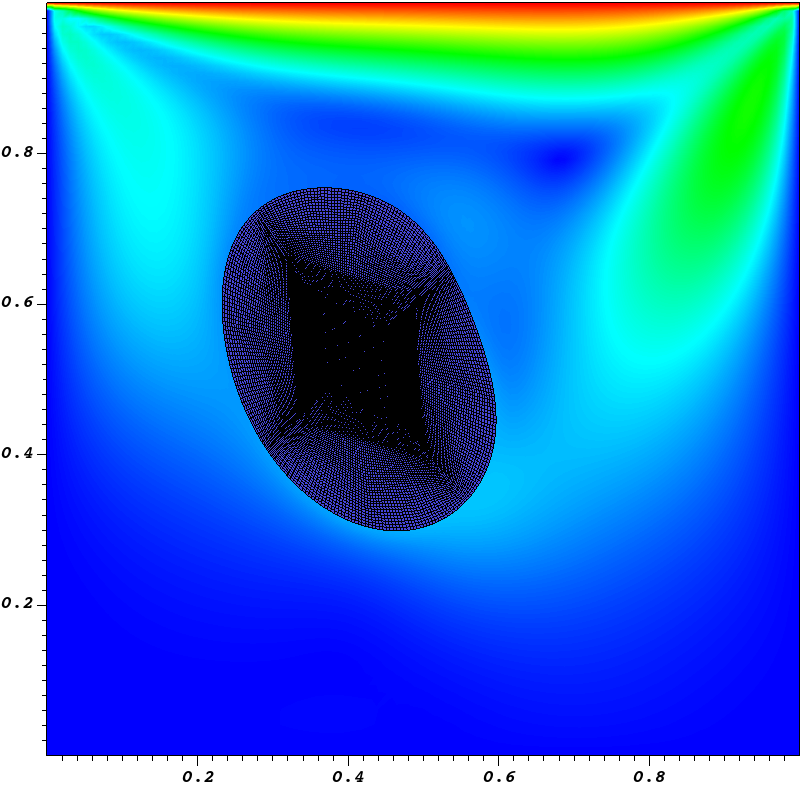}
			\caption{}
		\end{subfigure}
		\begin{subfigure}[b]{0.18\textwidth}
			\includegraphics[width=\textwidth]{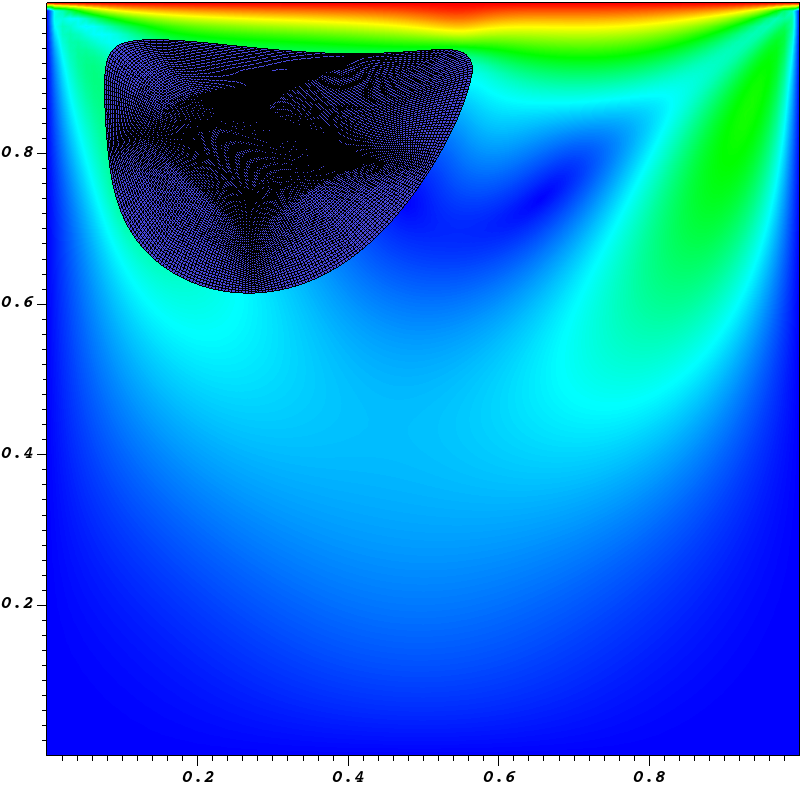}
			\caption{}
		\end{subfigure}
		\begin{subfigure}[b]{0.18\textwidth}
			\includegraphics[width=\textwidth]{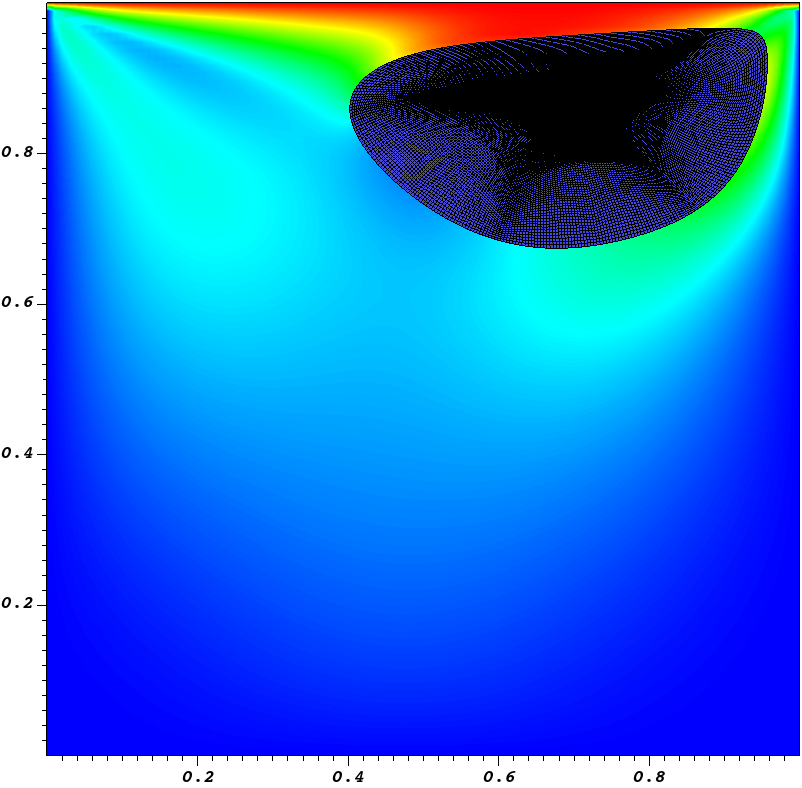}
			\caption{}
		\end{subfigure}
		\begin{subfigure}[b]{0.18\textwidth}
			\includegraphics[width=\textwidth]{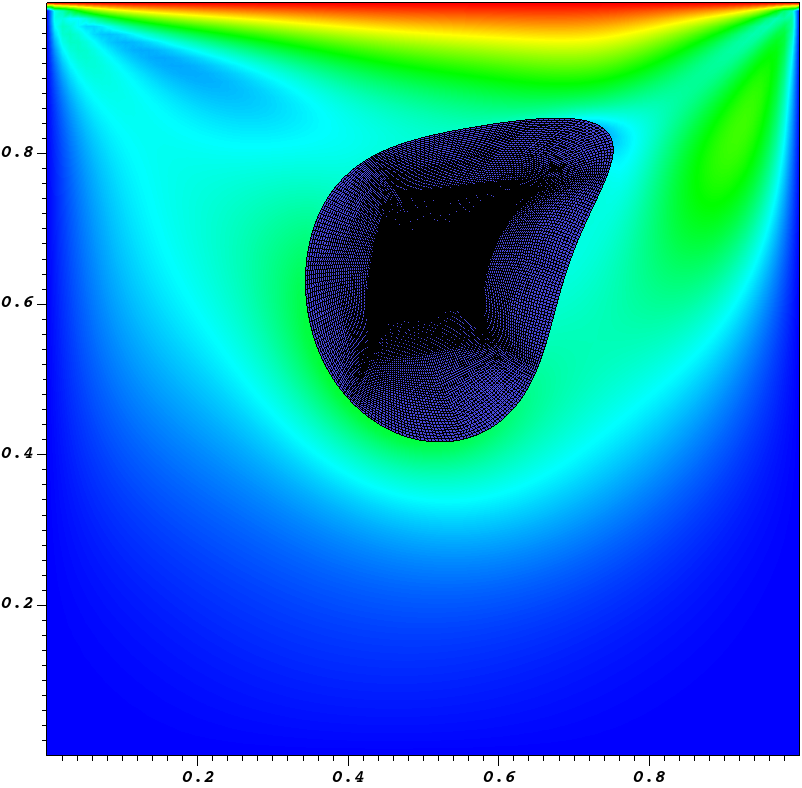}
			\caption{}
		\end{subfigure}
		
		\caption{The computed fluid velocity magnitude and motion of the soft disk in a lid-driven cavity. The top row corresponds to the unmodified IIM with $\mfac = 2$, while the bottom row shows results from the stabilized IIM with $\mfac = 0.25$ and $\epsilon = 9.21 \times 10^{-2}$, at times (a) and (f) $t = 0.0$ s, (b) and (g) $t = 2.0$ s, (c) and (h) $t = 4.0$ s, (d) and (i) $t = 6.0$ s, and (e) and (j) $t = 8.0$ s.}
		\label{fig:valve-side}
	\end{figure}
Next, we investigate the impact of the stabilized IIM in a flexible-body FSI model setup, in which the motion $\bxi(\bX, t)$ of the material interface $\Gamma_0$ is governed by the equations of flexible-body motion. For more technical details regarding the FSI coupling scheme, we refer to our previous work~\cite{KOLAHDOUZ2023112174}. Although we tested multiple flexible-body examples, the small-$\mfac$ instability observed in previous rigid-body cases did not appear in any of the flexible-body examples reported in our previous studies~\cite{KOLAHDOUZ2023112174}. Whether this instability arises in flexible-body problems remains an open question. Consequently, in these tests, we only investigate the impact of stabilization on the computed dynamics and not on the stability of the FSI algorithm.

This benchmark test considers a soft structure in a lid-driven cavity flow. The same benchmark has been used previously with the flexible-body ILE method~\cite{KOLAHDOUZ2023112174}, and the notation has been maintain the same for clarity. Notably, slightly different versions of this example have been explored in previous studies~\cite{Griffith_luo_2016,ROY20151167}. The computational domain is 
	\(\Omega = [0.0~\text{cm}, 1.0~\text{cm}] \times [0.0~\text{cm}, 1.0~\text{cm}]\), 
	a square of size \(L_x = L_y = 1.0~\text{cm}\). The immersed structure is initially a disk of radius 
	\(R = 0.2~\text{cm}\), centered at \((0.5~\text{cm}, 0.6~\text{cm})\). A uniform velocity 
	\(\mathbf{u} = ( 1~\text{cm}\cdot\text{s}^{-1},\, 0~\text{cm}\cdot\text{s}^{-1})\) 
	is imposed along the top boundary, and zero-velocity conditions are imposed at all other boundaries. 
	
	The fluid has a uniform density \(\rho_\text{f}  = 1.0~\text{g}\cdot\text{cm}^{-3}\) and dynamic viscosity 
	\(\mu_\text{f}  = 0.01~\text{g}\cdot(\text{cm}\cdot\text{s})^{-1}\). The soft structure has a shear modulus 
	\(G_\text{s}  = 0.1~\text{dyn}\cdot\text{cm}^{-2}\) and Poisson’s ratio \(\nu = 0.499\). The density ratio is  
	\(\rho_\text{s}/\rho_\text{f}\). 
	
	We consider the time span \(0.0~\text{s} \le t \le 10.0~\text{s}\), during which the disk goes through slightly more than one rotation inside the cavity. The Eulerian domain is discretized using a refinement ratio of 
	\(r = 2\) with \(N = 2\) Cartesian grid levels, with a grid spacing of 
	\(h_{\text{coarsest}} = L_x / 48\) on the coarsest level and 
	\(h_{\text{finest}} = L_x / 96\) on the finest level. The time step size is 
	\(\Delta t = (0.0005~\text{s}\cdot\text{cm}^{-1})\, h_{\text{finest}}\), 
	and the force penalty parameters are set to 
	\(\kappa = 1000\). 

Fig.~\ref{fig:valve-side} presents snapshots of the soft disk, including the corresponding fluid velocity magnitude and disk motion, at five different times for the unmodified IIM with $\mfac = 0.25$ and the stabilized IIM with $\mfac = 0.25$ and $\epsilon = 9.21\times10^{-2}$. The two methods exhibit almost identical dynamics under these settings. Fig.~\ref{fig:soft_disc} compares the trajectory of the centroid of the soft disk over time from the simulation results of the unmodified IIM with $\mfac = 2$ and the stabilized IIM with $\mfac = 0.25$ and $\epsilon = 0, 9.21\times10^{-6}, 9.21\times10^{-2}$. We observe excellent agreement in the centroid trajectory for a broad range of unmodified IIM with $\mfac = 2$ and the stabilized IIM with $\mfac = 0.25$ and $\epsilon = 0$ (which is equivalent to the unmodified IIM with $\mfac = 0.25$). No sign of small-$\mfac$ instability is observed. Moreover, we note  in Table~\ref{tabel_soft_disc} that small values of the stability coefficient ($\epsilon = 1.47\times10^{-6}$ and $1.47\times10^{-2}$) reproduce dynamics nearly identical to those of the unmodified IIM with both $\mfac = 0.25$ and $\mfac = 2$. 
}
\begin{figure}[hb!!]
	\centering
	\hskip -.3cm
	\includegraphics[width=.70\textwidth]{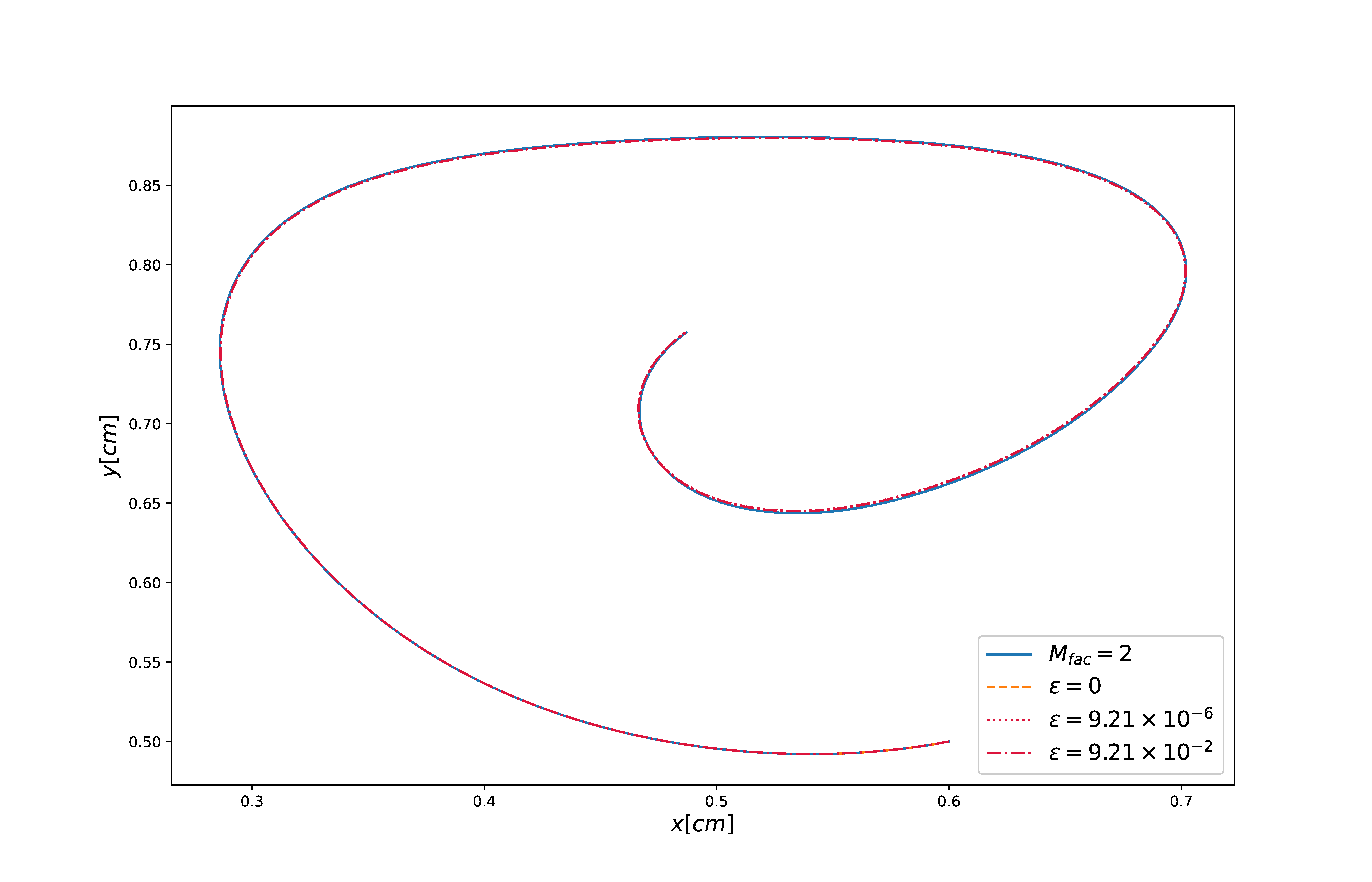}
	\caption{{\color{black} Trajectory of the centroid of the soft disk over time obtained using unmodified IIM with $\mfac=2$ and stabilized IIM with $\mfac=0.25$, $\epsilon=0,9.21\times10^{-6},9.21\times10^{-2}$. }}
	\label{fig:soft_disc}
\end{figure}

\begin{table}[ht!!]
	\centering
	\begin{tabular}{|l|l|l|}
		\hline
		& $\epsilon=9.21\times10^{-6}$ & $\epsilon=9.21\times10^{-2}$  \\ \hline
		Trajectory of the Centroid  Discrepancy  & $6.75\times10^{-3}$   & $2.37\times10^{-2}$  \\ \hline
	\end{tabular}
	\caption{\QS{Averaged RPD of trajectory of the centroid between the unmodified IIM~\cite{KOLAHDOUZ2020108854} with $\mfac = 0.25$ and the stabilized IIM with $\epsilon=1.024$ for $\mfac=0.25$.} }
	\label{tabel_soft_disc}
\end{table}
\section{Conclusions}
\label{sec:conclusion}
In this paper, \QS{ we introduced a stabilization scheme for our immersed interface method~\cite{KOLAHDOUZ2020108854} for discrete geometries that enable it to treat interface geometries with a broad range of element sizes. This scheme is intended to avoid instabilities that we observe when the unstabilized IIM is applied to} previously proposed {\color{black}immersed  interface method} small values of mesh factor ratio {\color{black}$\mfac = h_{\text{L}} / h_{\text{E}}$}, in which $h_{\text{L}}$ is the local Lagrangian element size and $h_{\text{E}}$ is the local Eulerian grid spacing. The proposed stabilized IIM demonstrates robust performance and stability across various test cases within the range $0.05 < \mfac < 1$, while maintaining dynamic behavior comparable to simulations performed using the unmodified IIM that satisfies the $\mfac > 1$ constraint. This approach enables a broader and more practical range of structure-to-fluid grid-size ratios without compromising accuracy. \QS{To demonstrate the effectiveness of our stabilization scheme, we consider problems with stationary interface configurations, as well as FSI models involving both rigid-body dynamics and elastodynamic structures models.} Numerical tests demonstrate that our stabilized formulation maintains accuracy comparable to our previously proposed IIM while allowing for much smaller or highly  disparate structure-to-fluid grid-size ratios previously considered infeasible. This advancement significantly broadens the applicability of the method to real-world FSI problems involving complex geometries and dynamic conditions, offering a robust and computationally efficient solution.

\QS{The choice of the stabilization parameter $\epsilon$ plays a crucial role in ensuring both numerical stability and physical fidelity. A general guideline is to select a value large enough to prevent numerical instability but small enough to avoid artificially altering the system's physical dynamics. While our tests on flexible bodies did not exhibit the instabilities seen in rigid-body cases, this suggests that elasticity may help mitigate the need for stabilization. Whether this instability can ever arise in flexible-body problems remains an open question. There are also many open problems to explore, including a more rigorous investigation into the stabilizing properties of structural elasticity and the development of a more locally and geometrically tailored stabilization scheme.}
\section{Acknowledgements}
We gratefully acknowledge research support through NIH Awards HL143336 and HL157631 and  NSF Awards CBET 1757193, OAC 1652541, and  OAC 1931516. Computations were performed using facilities provided by University of
North Carolina at Chapel Hill through the Research Computing division of UNC Information Technology Services. We also thank Dr. Cole Gruninger for his valuable suggestions on the manuscript.

{\color{black}This manuscript is the result of funding in whole or in part by the National Institutes of Health (NIH). It is subject to the NIH Public Access Policy. Through acceptance of this federal funding, NIH has been given a right to make this manuscript publicly available in PubMed Central upon the Official Date of Publication, as defined by NIH.}
\bibliographystyle{ieeetr}
\bibliography{main}

\begin{thebibliography}{10}

\bibitem{van_2020}
E.~van~der Giessen, P.~A. Schultz, N.~Bertin, V.~V. Bulatov, W.~Cai,
  G.~Csányi, S.~M. Foiles, M.~G.~D. Geers, C.~González, M.~Hütter, W.~K.
  Kim, D.~M. Kochmann, J.~LLorca, A.~E. Mattsson, J.~Rottler, A.~Shluger, R.~B.
  Sills, I.~Steinbach, A.~Strachan, and E.~B. Tadmor, ``Roadmap on multiscale
  materials modeling,'' {\em Modelling and Simulation in Materials Science and
  Engineering}, vol.~28, p.~43001, Mar 2020.

\bibitem{jne2010005}
M.~Avramova, A.~Abarca, J.~Hou, and K.~Ivanov, ``Innovations in multi-physics
  methods development, validation, and uncertainty quantification,'' {\em
  Journal of Nuclear Engineering}, vol.~2, no.~1, pp.~44--56, 2021.

\bibitem{Bause2025SystematicHI}
O.~Bause, J.~Werner, and O.~Bringmann, ``Systematic hardware integration
  testing for smart video-based medical device prototypes,'' {\em ArXiv},
  vol.~abs/2504.19533, 2025.

\bibitem{Donea2004}
J.~Donea, A.~Huerta, J.-P. Ponthot, and A.~Rodríguez-Ferran, {\em Arbitrary
  Lagrangian–Eulerian Methods}, ch.~14.
\newblock John Wiley \& Sons Ltd., 2004.

\bibitem{ADJERID2015170}
S.~Adjerid, N.~Chaabane, and T.~Lin, ``An immersed discontinuous finite element
  method for {Stokes} interface problems,'' {\em Computer Methods in Applied
  Mechanics and Engineering}, vol.~293, pp.~170--190, 2015.

\bibitem{OlshanskiiQiQuaini2021}
M.~Olshanskii, A.~Quaini, and Q.~Sun, ``A finite element method for two-phase
  flow with material viscous interface,'' {\em Computational Methods in Applied
  Mathematics}, vol.~22, pp.~443--464, 2021.

\bibitem{OlshanskiiQiQuaini2022}
M.~Olshanskii, A.~Quaini, and Q.~Sun, ``An unfitted finite element method for
  two-phase {Stokes} problems with slip between phases,'' {\em Journal of
  Scientific Computing}, vol.~89, 2021.

\bibitem{PESKIN1972252}
C.~S. Peskin, ``Flow patterns around heart valves: A numerical method,'' {\em
  Journal of Computational Physics}, vol.~10, no.~2, pp.~252--271, 1972.

\bibitem{Peskin_2002}
C.~S. Peskin, ``The immersed boundary method,'' {\em Acta Numerica}, vol.~11,
  p.~479–517, 2002.

\bibitem{GRIFFITH200575}
B.~E. Griffith and C.~S. Peskin, ``On the order of accuracy of the immersed
  boundary method: Higher order convergence rates for sufficiently smooth
  problems,'' {\em Journal of Computational Physics}, vol.~208, no.~1,
  pp.~75--105, 2005.

\bibitem{FAI2018319}
T.~G. Fai and C.~H. Rycroft, ``Lubricated immersed boundary method in two
  dimensions,'' {\em Journal of Computational Physics}, vol.~356, pp.~319--339,
  2018.

\bibitem{leveque1994immersed}
R.~J. Leveque and Z.~Li, ``The immersed interface method for elliptic equations
  with discontinuous coefficients and singular sources,'' {\em SIAM Journal on
  Numerical Analysis}, vol.~31, no.~4, pp.~1019--1044, 1994.

\bibitem{LeVeque97}
R.~J. LeVeque and Z.~Li, ``Immersed interface methods for stokes flow with
  elastic boundaries or surface tension,'' {\em SIAM Journal on Scientific
  Computing}, vol.~18, no.~3, pp.~709--735, 1997.

\bibitem{LeVequeRandall2003}
L.~Lee and R.~J. LeVeque, ``An immersed interface method for incompressible
  {Navier--Stokes} equations,'' {\em SIAM Journal on Scientific Computing},
  vol.~25, no.~3, pp.~832--856, 2003.

\bibitem{KOLAHDOUZ2020108854}
E.~M. Kolahdouz, A.~P.~S. Bhalla, B.~A. Craven, and B.~E. Griffith, ``An
  immersed interface method for discrete surfaces,'' {\em Journal of
  Computational Physics}, vol.~400, p.~108854, 2020.

\bibitem{KOLAHDOUZ2021110442}
E.~M. Kolahdouz, A.~P.~S. Bhalla, L.~N. Scotten, B.~A. Craven, and B.~E.
  Griffith, ``A sharp interface {Lagrangian-Eulerian} method for rigid-body
  fluid-structure interaction,'' {\em Journal of Computational Physics},
  vol.~443, p.~110442, 2021.

\bibitem{KOLAHDOUZ2023112174}
E.~M. Kolahdouz, D.~R. Wells, S.~Rossi, K.~I. Aycock, B.~A. Craven, and B.~E.
  Griffith, ``A sharp interface {Lagrangian--Eulerian} method for flexible-body
  fluid-structure interaction,'' {\em Journal of Computational Physics},
  vol.~488, p.~112174, 2023.

\bibitem{xusheng06}
S.~Xu and Z.~J. Wang, ``Systematic derivation of jump conditions for the
  immersed interface method in three-dimensional flow simulation,'' {\em SIAM
  Journal on Scientific Computing}, vol.~27, no.~6, pp.~1948--1980, 2006.

\bibitem{Lai2001ARO}
M.-C. Lai and Z.~Li, ``A remark on jump conditions for the three-dimensional
  {Navier--Stokes} equations involving an immersed moving membrane,'' {\em
  Applied Mathematics Letters}, vol.~14, pp.~149--154, 2001.

\bibitem{tikhonov1977solutions}
A.~N. Tikhonov and V.~Y. Arsenin, {\em Solutions of ill-posed problems}.
\newblock Washington, D.C.: John Wiley \& Sons, New York: V. H. Winston \&
  Sons, 1977.
\newblock Translated from the Russian, Preface by translation editor Fritz
  John, Scripta Series in Mathematics.

\bibitem{GOLDSTEIN1993354}
D.~Goldstein, R.~Handler, and L.~Sirovich, ``Modeling a no-slip flow boundary
  with an external force field,'' {\em Journal of Computational Physics},
  vol.~105, no.~2, pp.~354--366, 1993.

\bibitem{GRIFFITH20097565}
B.~E. Griffith, ``An accurate and efficient method for the incompressible
  {Navier--Stokes} equations using the projection method as a preconditioner,''
  {\em Journal of Computational Physics}, vol.~228, no.~20, pp.~7565--7595,
  2009.

\bibitem{Boyce052011}
B.~E. Griffith, ``Immersed boundary model of aortic heart valve dynamics with
  physiological driving and loading conditions,'' {\em International Journal
  for Numerical Methods in Biomedical Engineering}, vol.~28, no.~3,
  pp.~317--345, 2012.

\bibitem{tan2009}
Z.~Tan, D.~Le, K.~Lim, and B.~Khoo, ``An immersed interface method for the
  incompressible {Navier--Stokes} equations with discontinuous viscosity across
  the interface,'' {\em SIAM Journal on Scientific Computing}, vol.~31, no.~3,
  pp.~1798--1819, 2009.

\bibitem{Dongwang}
D.~Wang and S.~J. Ruuth, ``Variable step-size implicit-explicit linear
  multistep methods for time-dependent partial differential equations,'' {\em
  Journal of Computational Mathematics}, vol.~26, no.~6, pp.~838--855, 2008.

\bibitem{Griffith_2012}
B.~E. Griffith, ``On the volume conservation of the immersed boundary method,''
  {\em Communications in Computational Physics}, vol.~12, no.~2, p.~401–432,
  2012.

\bibitem{Griffith_luo_2016}
B.~E. Griffith and X.~Y. Luo, ``Hybrid finite difference/finite element
  immersed boundary method,'' {\em International Journal for Numerical Methods
  in Biomedical Engineering}, vol.~33, no.~12, p.~e2888, 2017.
\newblock e2888 cnm.2888.

\bibitem{TAIRA20072118}
K.~Taira and T.~Colonius, ``The immersed boundary method: A projection
  approach,'' {\em Journal of Computational Physics}, vol.~225, no.~2,
  pp.~2118--2137, 2007.

\bibitem{Braza_Chassaing_Minh_1986}
M.~Braza, P.~Chassaing, and H.~H. Minh, ``Numerical study and physical analysis
  of the pressure and velocity fields in the near wake of a circular
  cylinder,'' {\em Journal of Fluid Mechanics}, vol.~165, p.~79–130, 1986.

\bibitem{LIU199835}
C.~Liu, X.~Zheng, and C.~Sung, ``Preconditioned multigrid methods for unsteady
  incompressible flows,'' {\em Journal of Computational Physics}, vol.~139,
  no.~1, pp.~35--57, 1998.

\bibitem{XU20082068}
S.~Xu and Z.~J. Wang, ``A 3d immersed interface method for fluid–solid
  interaction,'' {\em Computer Methods in Applied Mechanics and Engineering},
  vol.~197, no.~25, pp.~2068--2086, 2008.

\bibitem{Fornberg_1988}
B.~Fornberg, ``Steady viscous flow past a sphere at high {Reynolds} numbers,''
  {\em Journal of Fluid Mechanics}, vol.~190, p.~471–489, 1988.

\bibitem{TURTON198683}
R.~Turton and O.~Levenspiel, ``A short note on the drag correlation for
  spheres,'' {\em Powder Technology}, vol.~47, no.~1, pp.~83--86, 1986.

\bibitem{FADLUN200035}
E.~Fadlun, R.~Verzicco, P.~Orlandi, and J.~Mohd-Yusof, ``Combined
  immersed-boundary finite-difference methods for three-dimensional complex
  flow simulations,'' {\em Journal of Computational Physics}, vol.~161, no.~1,
  pp.~35--60, 2000.

\bibitem{Campregher09}
R.~Campregher, J.~Militzer, S.~Mansur, and A.~Neto, ``Computations of the flow
  past a still sphere at moderate {Reynolds} numbers using an immersed boundary
  method,'' {\em Journal of The Brazilian Society of Mechanical Sciences and
  Engineering - J BRAZ SOC MECH SCI ENG}, vol.~31, 2009.

\bibitem{AHN2006671}
H.~T. Ahn and Y.~Kallinderis, ``Strongly coupled flow/structure interactions
  with a geometrically conservative ale scheme on general hybrid meshes,'' {\em
  Journal of Computational Physics}, vol.~219, no.~2, pp.~671--696, 2006.

\bibitem{Borazjani2008}
I.~Borazjani, L.~Ge, and F.~Sotiropoulos, ``Curvilinear immersed boundary
  method for simulating fluid structure interaction with complex 3d rigid
  bodies,'' {\em Journal of Computational Physics}, vol.~227, pp.~7587--7620,
  2008.

\bibitem{Bao2012}
Y.~Bao, C.~Huang, D.~Zhou, J.~Tu, and Z.~Han, ``Two-degree-of-freedom
  flow-induced vibrations on isolated and tandem cylinders with varying natural
  frequency ratios,'' {\em Journal of Fluids and Structures}, vol.~35,
  p.~50–75, 2012.

\bibitem{Jianming2008}
J.~Yang, S.~Preidikman, and E.~Balaras, ``A strongly coupled, embedded-boundary
  method for fluid–structure interactions of elastically mounted rigid
  bodies,'' {\em Journal of Fluids and Structures}, vol.~24, pp.~167--182,
  2008.

\bibitem{Jianming2015}
J.~Yang and F.~Stern, ``A non-iterative direct forcing immersed boundary method
  for strongly-coupled fluid–solid interactions,'' {\em Journal of
  Computational Physics}, vol.~295, pp.~779--804, 2015.

\bibitem{KIM2018296}
W.~Kim, I.~Lee, and H.~Choi, ``A weak-coupling immersed boundary method for
  fluid–structure interaction with low density ratio of solid to fluid,''
  {\em Journal of Computational Physics}, vol.~359, pp.~296--311, 2018.

\bibitem{Blackburn1993TwoAT}
H.~Blackburn and G.~Karniadakis, ``Two-and three-dimensional simulations of
  vortex-induced vibration or a circular cylinder,'' vol.~All Days of {\em
  International Ocean and Polar Engineering Conference}, pp.~ISOPE--I--93--317,
  1993.

\bibitem{Liuhu18}
C.~{Liu} and C.~{Hu}, ``{Block-based adaptive mesh refinement for
  fluid-structure interactions in incompressible flows},'' {\em Computer
  Physics Communications}, vol.~232, pp.~104--123, 2018.

\bibitem{Feng_Hu_Joseph_1994}
J.~Feng, H.~H. Hu, and D.~D. Joseph, ``Direct simulation of initial value
  problems for the motion of solid bodies in a newtonian fluid part 1.
  sedimentation,'' {\em Journal of Fluid Mechanics}, vol.~261, p.~95–134,
  1994.

\bibitem{LACIS2016300}
U.~Lācis, K.~Taira, and S.~Bagheri, ``A stable fluid–structure-interaction
  solver for low-density rigid bodies using the immersed boundary projection
  method,'' {\em Journal of Computational Physics}, vol.~305, pp.~300--318,
  2016.

\bibitem{BHALLA2013446}
A.~P.~S. Bhalla, R.~Bale, B.~E. Griffith, and N.~A. Patankar, ``A unified
  mathematical framework and an adaptive numerical method for fluid–structure
  interaction with rigid, deforming, and elastic bodies,'' {\em Journal of
  Computational Physics}, vol.~250, pp.~446--476, 2013.

\bibitem{ROY20151167}
S.~Roy, L.~Heltai, and F.~Costanzo, ``Benchmarking the immersed finite element
  method for fluid–structure interaction problems,'' {\em Computers \&
  Mathematics with Applications}, vol.~69, no.~10, pp.~1167--1188, 2015.

\end{thebibliography}
\end{document}